\documentclass[sigconf]{acmart}

\usepackage{comment}
\usepackage{balance}
\usepackage{graphicx} 
\usepackage{subcaption}
\usepackage{soul}
\usepackage{tcolorbox}
\usepackage{balance}
\usepackage{stfloats}
\usepackage{booktabs}
\usepackage{makecell}
\usepackage{caption}
\usepackage{amsmath}
\usepackage{graphicx} 
\usepackage{enumitem}
\usepackage{gensymb}
\usepackage{multirow}
\usepackage{algorithm}
\usepackage{algorithmic}
\usepackage{pifont}
\usepackage{cleveref}

\usepackage{url}

\newcommand{\blue}[1]{\textcolor{black}{#1}}

\AtBeginDocument{%
  }

\setcopyright{acmlicensed}
\copyrightyear{2026}
\acmYear{2026}
\acmDOI{XXXXXXX.XXXXXXX}
\acmConference[CCS '26]{ACM SIGSAC Conference on Computer and Communications Security}{November 15-19, 2026}{Hague, Netherlands}
\acmISBN{978-1-4503-XXXX-X/2026/06}

\renewcommand\footnotetextcopyrightpermission[1]{}
\begin{document}

\title{Snatcher: Apple Find My Network Exposes Your Lost Devices To Strangers}

\title{Snatcher: Apple Find My Network Exposes Your Lost Devices To Strangers}

\renewcommand{\shortauthors}{Ren, Zhang, Liu, and Li}

\author{Zhenyu Ren$^\ast$, Yanbo Zhang$^\ast$, Boya Liu, Mo Li$^\dagger$}

\thanks{$^\ast$Both authors contributed equally to this research.}
\thanks{$^\dagger$Corresponding author.}

\affiliation{%
  \institution{The Hong Kong University of Science and Technology}
  \city{Hong Kong}
  \country{China}
}

\email{{zrenax, bliuby}@connect.ust.hk, {yanbozhang, lim}@ust.hk}

\renewcommand{\shortauthors}{Ren et al.}

\begin{abstract}
Apple’s Find My network connects nearly one billion devices to locate missing property via Bluetooth Low Energy (BLE). 
\blue {This paper reveals that insecure BLE advertisements and design tradeoffs allow unauthorized discovery and physical theft of lost Apple devices.} 
We develop \textit{Snatcher}, an attack and analysis framework implemented fully on Android smartphones without specialized hardware. 
\textit{Snatcher} identifies vulnerabilities in unencrypted BLE advertisements, unauthenticated acoustic triggers, and slow MAC address randomization.
Through three levels---sound-based direction finding, RSSI--IMU sensor-fusion navigation, and spatial-temporal clustering---our Android-based platform physically tracks and locates lost Apple accessories and devices in real-world tests. 
Our results highlight a crucial conflict between privacy protection, anti-stalking design, and physical security, urging Apple to strengthen Find My defenses.
\end{abstract}

\begin{CCSXML}
<ccs2012>
   <concept>
       <concept_id>10002978.10003014.10003017</concept_id>
       <concept_desc>Security and privacy~Mobile and wireless security</concept_desc>
       <concept_significance>500</concept_significance>
       </concept>
 </ccs2012>
\end{CCSXML}

\ccsdesc[500]{Security and privacy~Mobile and wireless security}

\keywords{Find My network; Physical Security; Bluetooth Low Energy (BLE)}

\received{20 February 2007}
\received[revised]{12 March 2009}
\received[accepted]{5 June 2009}

\maketitle

\section{Introduction}
\label{sec:introduction}

Offline finding networks, large-scale, crowd-assisted ecosystems for locating lost property, have become a mainstream technology adopted by major vendors, including Apple, Google, and Samsung~\cite{findmy, AppleFindMySpec2020, bottger2025okay, google_fmdn_fhn_spec_v1_3, samsung_smartthings_find_dev_guide, yu2024security}. Apple's Find My network, launched in 2019, is a prominent example, harnessing a global architecture of nearly one billion devices. \blue{When an item is lost, it is detected by nearby participating devices, which then securely relay its location to the owner through an end-to-end encrypted channel~\cite{AppleFindMySpec2020}.} This design provides a powerful recovery tool while preserving the confidentiality of the location data.

However, the design choices that make these networks effective for owners simultaneously introduce critical vulnerabilities that expose lost devices to physical theft. 
\blue{In this paper, we identify a tripartite of \emph{structural vulnerabilities} stemming from the inherent tension between usability, anti-stalking measures, and physical security.}
First, to ensure universal discoverability, the network broadcasts the \textit{existence} of a lost device via unencrypted Bluetooth Low Energy (BLE) advertisements (\textit{Vulnerability I: Cleartext Discovery Beacon}). While location reports are encrypted, these cleartext beacons allow any nearby adversary to instantly detect and identify a high-value lost item without authorization, transforming it from a needle in a haystack into a glaring target. Second, the system's primary anti-stalking defense creates an exploitable acoustic side-channel (\textit{Vulnerability II: Unauthenticated Actuation}). The network permits any non-owner to connect to lost accessories (e.g., AirTags) and trigger a sound~\cite{10.1145/3507657.3528546, apple_airtag_unwanted_tracking_2022, google_unknown_tracker_alerts_2023, AppleFindMySpec2020}. This feature, intended to help victims find unwanted trackers, is weaponized by an attacker to pinpoint the exact location of a valuable item through auditory feedback. Finally, the mechanism for privacy protection, MAC address randomization, suffers from excessively long update intervals (\textit{Vulnerability III: Extended Identity Persistence})~\cite{AppleFindMySpec2020, eldridge2024abuse, Martin2019HandoffAY}. Our measurements reveal that rotation periods—up to 24 hours for AirTags—provide a vast window of opportunity for an attacker. This prolonged identity persistence allows for continuous signal-based tracking and navigation, enabling them to physically approach and steal the device long before its digital identifier changes.

\begin{figure*}[htb]
    \centering
    \includegraphics[width=\linewidth]{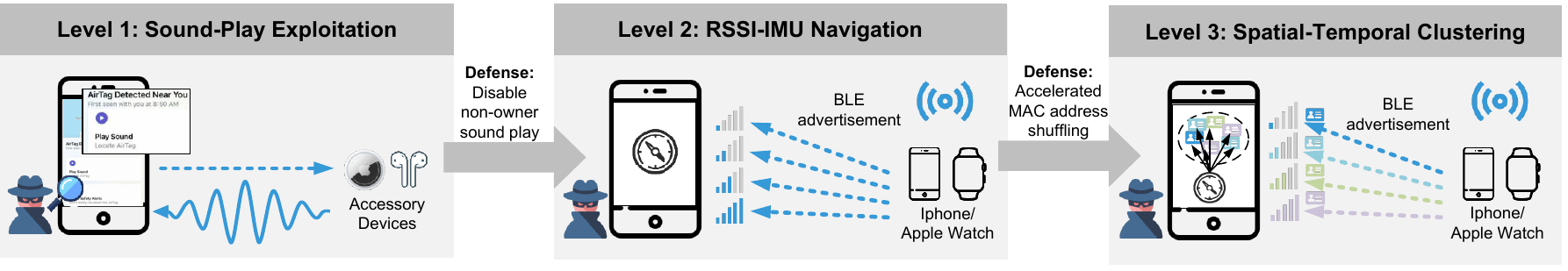}
    \caption{\blue{\textit{Snatcher} investigates the vulnerabilities of Apple Find My network and demonstrates a three-level attack model.}}
    \label{fig:scenario}
\end{figure*}

To systematically exploit these vulnerabilities, we introduce \textit{Snatcher}, a progressive, three-level physical attack framework designed from an adversary's perspective. Each level addresses a distinct set of device capabilities and potential countermeasures, demonstrating a clear escalation in attack sophistication (as illustrated in Figure~\ref{fig:scenario}).

\textit{Level 1: Acoustic Direction Finding.} The first level targets accessories like AirTags and AirPods, which are particularly vulnerable due to their exploitable \textit{non-owner sound play} feature. By programmatically triggering the device's audible alert, an attacker can transform a hidden digital signal into a physically perceptible one. We design an \textit{Acoustic Direction Finding} algorithm that leverages a standard smartphone's microphone to compute a reliable heading, enabling precise navigation toward the sound source without specialized hardware.

\textit{Level 2: RSSI-IMU Navigation.} The second level targets more resilient devices, such as iPhones and Apple Watches, which are immune to the acoustic attack. To overcome this, we pivot to a purely passive attack that exploits their continuous BLE advertising behavior. We develop a novel \textit{RSSI-IMU Navigation} algorithm that fuses the noisy Received Signal Strength Indicator (RSSI) data with the attacker's own motion data from the Inertial Measurement Unit (IMU). This sensor-fusion approach enables the attacker to perform a physical gradient ascent, inferring the target's direction from signal trends even in the absence of any acoustic feedback.

\textit{Level 3: De-anonymization via Spatial-Temporal Clustering.} The final level confronts a more advanced defense: high-frequency MAC address randomization. This countermeasure aims to thwart tracking by constantly changing the device's identifier. To defeat this, we propose a \textit{Spatial-Temporal Clustering} strategy. This algorithm de-anonymizes the target by correlating signal characteristics (RSSI) with their spatial-temporal context. It effectively stitches together the fragmented identities observed over time, allowing the attacker to maintain a persistent lock on a single physical device amidst a chaotic stream of rotating addresses.

In summary, our main contributions are:
\begin{itemize}
    \item \blue{\textbf{Vulnerability Analysis.} We perform the first systematic analysis of the physical security risks in Apple's Find My network.} 
    \blue{We identify and demonstrate three critical \emph{design tradeoffs}: an un-obfuscated discovery beacon, an exploitable acoustic side-channel, and insufficient MAC address randomization, all of which facilitate device theft.}
    \item \textbf{Progressive Attack Framework.} We design and formalize \textit{Snatcher}, a three-level attack framework that progressively escalates in sophistication. It includes: (1) \textit{Acoustic Direction Finding} for vulnerable accessories, (2) \textit{RSSI-IMU Navigation} for silent devices, and (3) \textit{Spatial-Temporal Clustering} to defeat advanced randomization defenses.
    \item \textbf{Real-World Implementation and Evaluation.} We implement a full prototype of \textit{Snatcher} on commodity Android smartphones, requiring no specialized hardware. Through extensive real-world experiments, we validate the effectiveness of our attacks across various Apple devices and environments, demonstrating a tangible and significant threat.
\end{itemize}

The remainder of this paper is organized as follows. We provide background on the Find My network in Section~\ref{sec:primer} and present our threat model and vulnerability analysis in Section~\ref{sec:threat_model}. We detail the design of our three-level attack in Section~\ref{sec:design}. The implementation and evaluation are presented in Section~\ref{sec:evaluation}. Finally, we discuss related work in Section~\ref{sec:relatedwork} and conclude in Section~\ref{sec:conclusion}.

\begin{figure}[t]
    \centering
    \includegraphics[width=\linewidth]{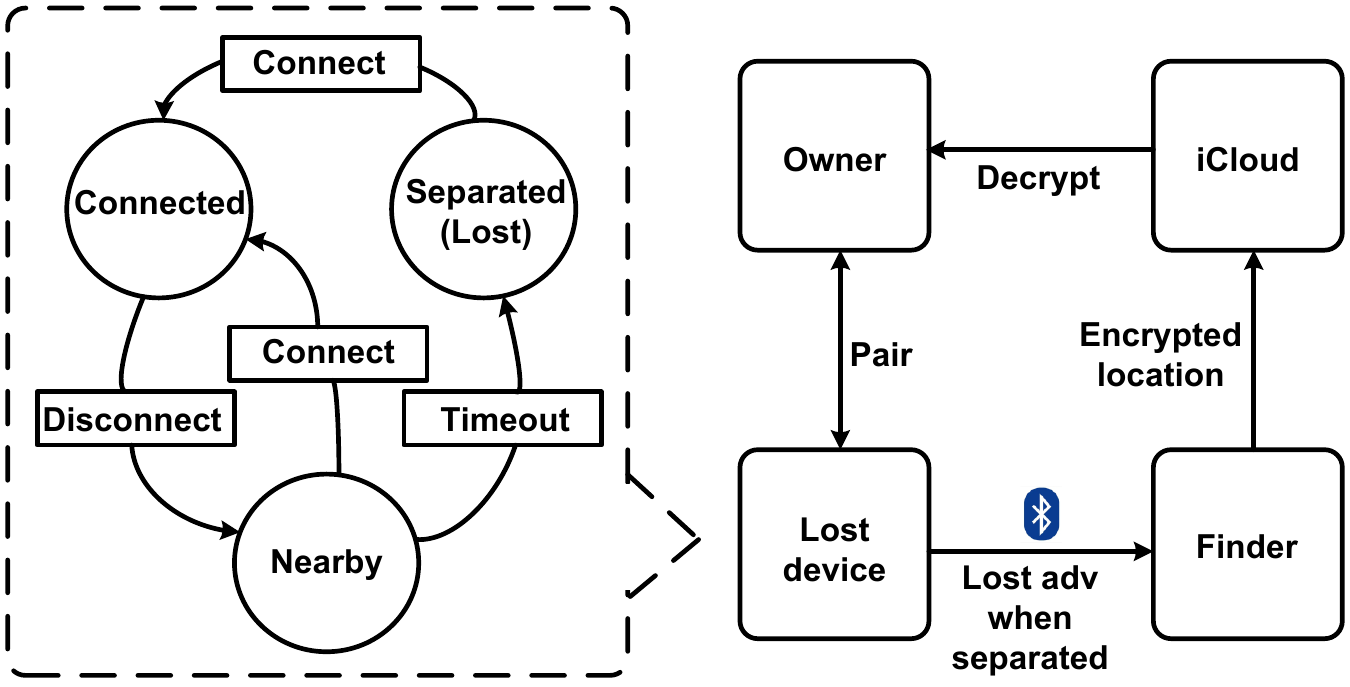}
    \caption{\blue{Operation flow of Apple Find My network.}}
    \label{fig:findmy}
\end{figure}

\vspace{-1pt}
\section{Primer on Apple Find My Network}
\label{sec:primer}
\subsection{Find My Network Operation Flow}
\label{subsec:operation_flow}

The Find My network operates as a distributed state machine that governs how devices transition between connected and lost modes, as illustrated in Figure~\ref{fig:findmy}. Each device can reside in one of three primary states -- \texttt{Connected}, \texttt{Nearby}, and \texttt{Separated (Lost)} — depending on its Bluetooth connectivity and proximity to the owner's primary device \cite{AppleFindMySpec2020}.

In the \texttt{Connected} state, the accessory has an active and encrypted Bluetooth connection with its paired owner device (like an iPhone). An accessory enters this state immediately after Find My network pairing is complete or whenever the owner's device reconnects.

When the device moves out of range or temporarily loses the Bluetooth connection, it enters the \texttt{Nearby} state. 
During the \texttt{Nearby} state, the device periodically attempts to re-establish the connection. This retry behavior allows transient disconnections, such as those caused by environmental interference, to be recovered without triggering an unnecessary lost event. In both the \texttt{Nearby} and \texttt{Connected} states, the device periodically advertises Find My BLE advertisement frames without a public key payload to indicate its presence and avoid being detected by other devices.

If reconnection attempts fail after a predefined timeout duration, the device transitions into the \texttt{Separated (Lost)} state. 
In this state, the device begins advertising specialized Find My BLE advertising packets containing a public key associated with its Find My identity. Surrounding Apple devices, acting as \texttt{Finders}, passively scan for these advertisements. Upon detecting a advertising packet, a Finder captures its own location, encrypts it with the public key from the advertisement to create a location report, and uploads the report to Apple's iCloud server. Only the owner can later decrypt these reports using their private key, ensuring end-to-end confidentiality of location data \cite{apple_platform_security_2024}. Moreover, for \textit{Anti-Stalking} purposes, accessories in the \texttt{Separated} state also allow non-owners to trigger a sound, as detailed in Subsection~\ref{subsec:non_owner_sound}.

\begin{figure}[t]
    \centering
    \begin{subfigure}[t]{0.24\linewidth}
        \centering
        \includegraphics[width=\linewidth]{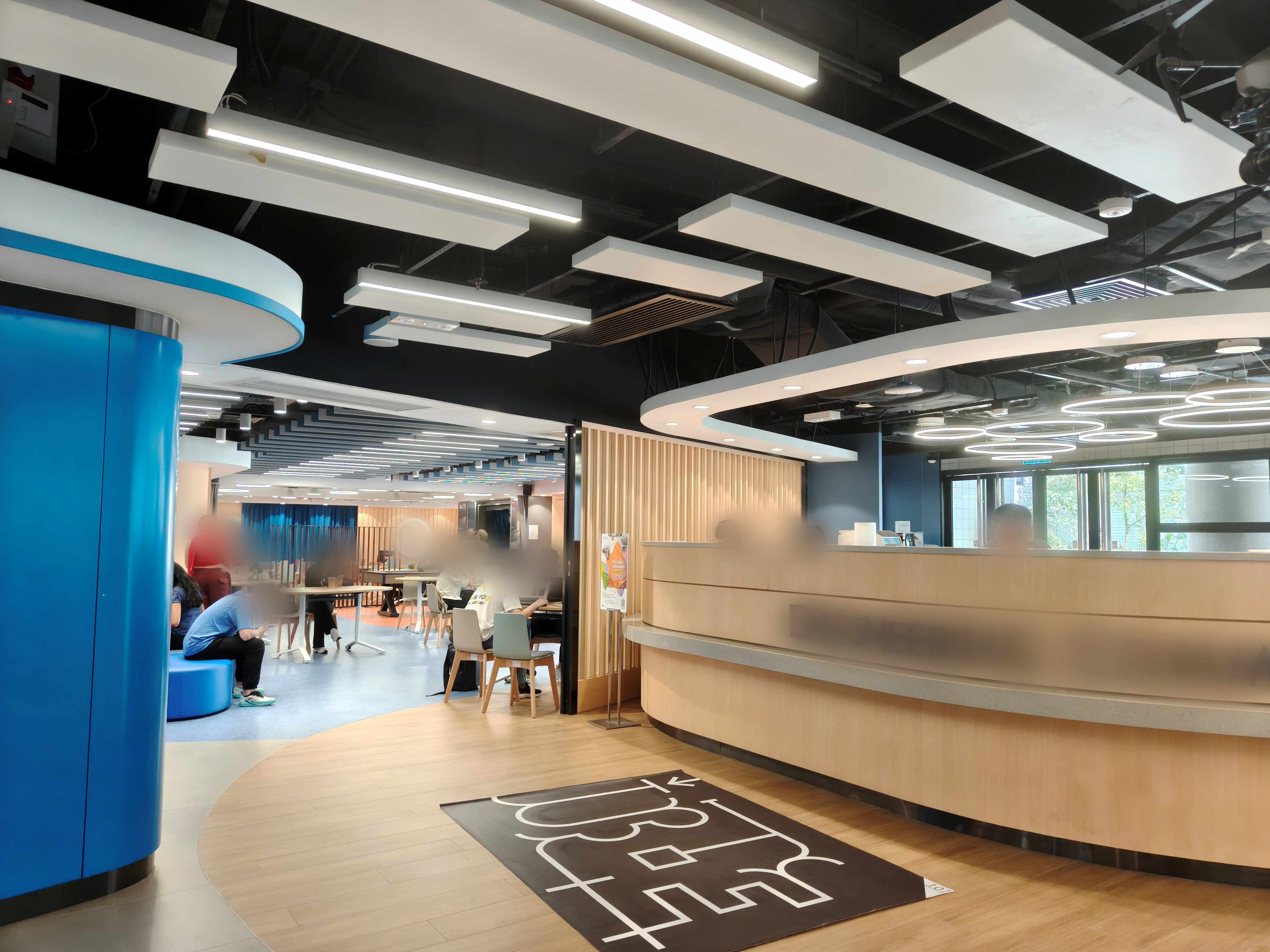}
        \caption{\blue{Office}}
        \label{fig:office}
    \end{subfigure}
    \hfill
    \begin{subfigure}[t]{0.24\linewidth}
        \centering
        \includegraphics[width=\linewidth]{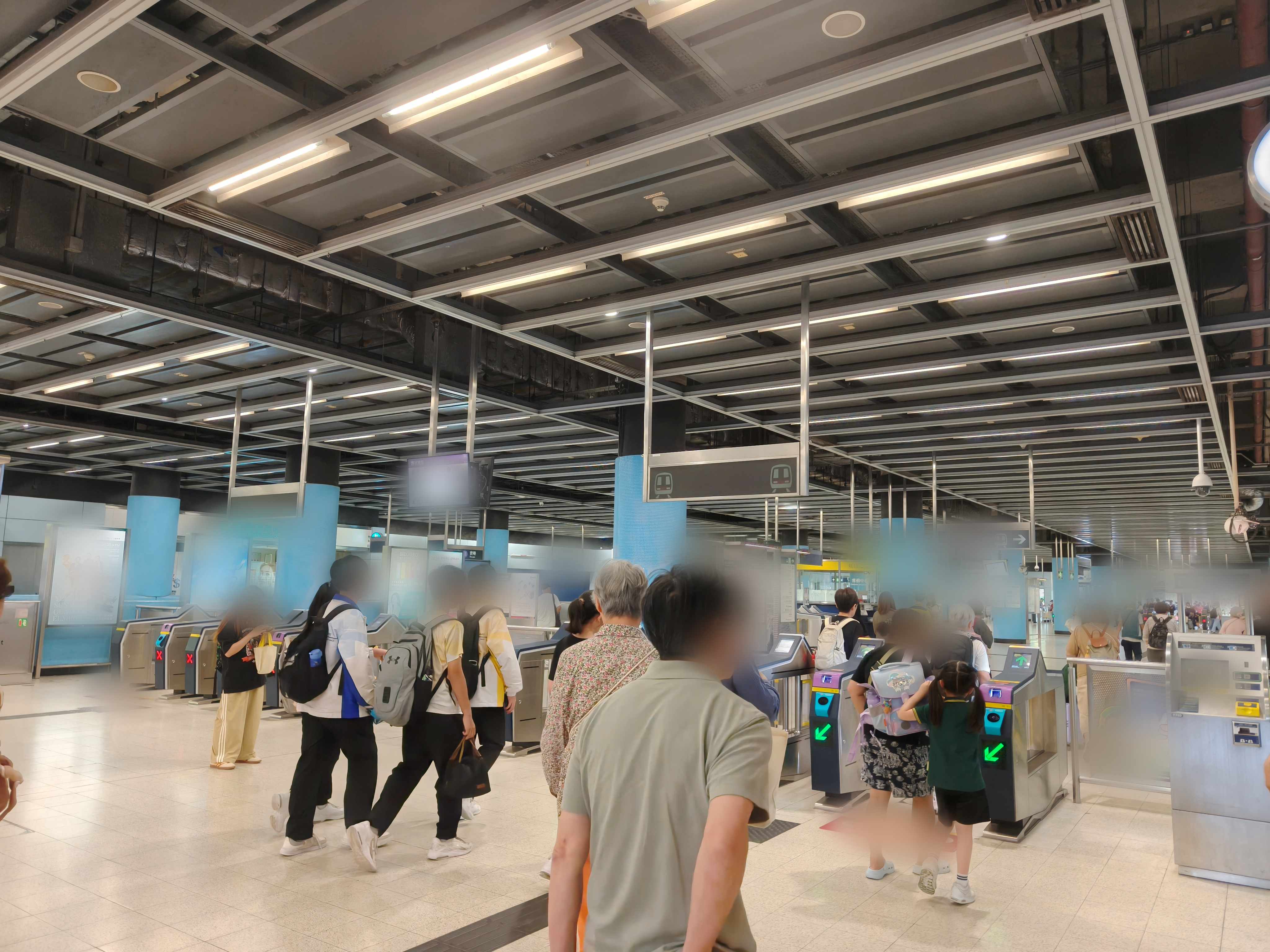}
        \caption{\blue{Subway}}
        \label{fig:subway}
    \end{subfigure}
    \hfill
    \begin{subfigure}[t]{0.24\linewidth}
        \centering
        \includegraphics[width=\linewidth]{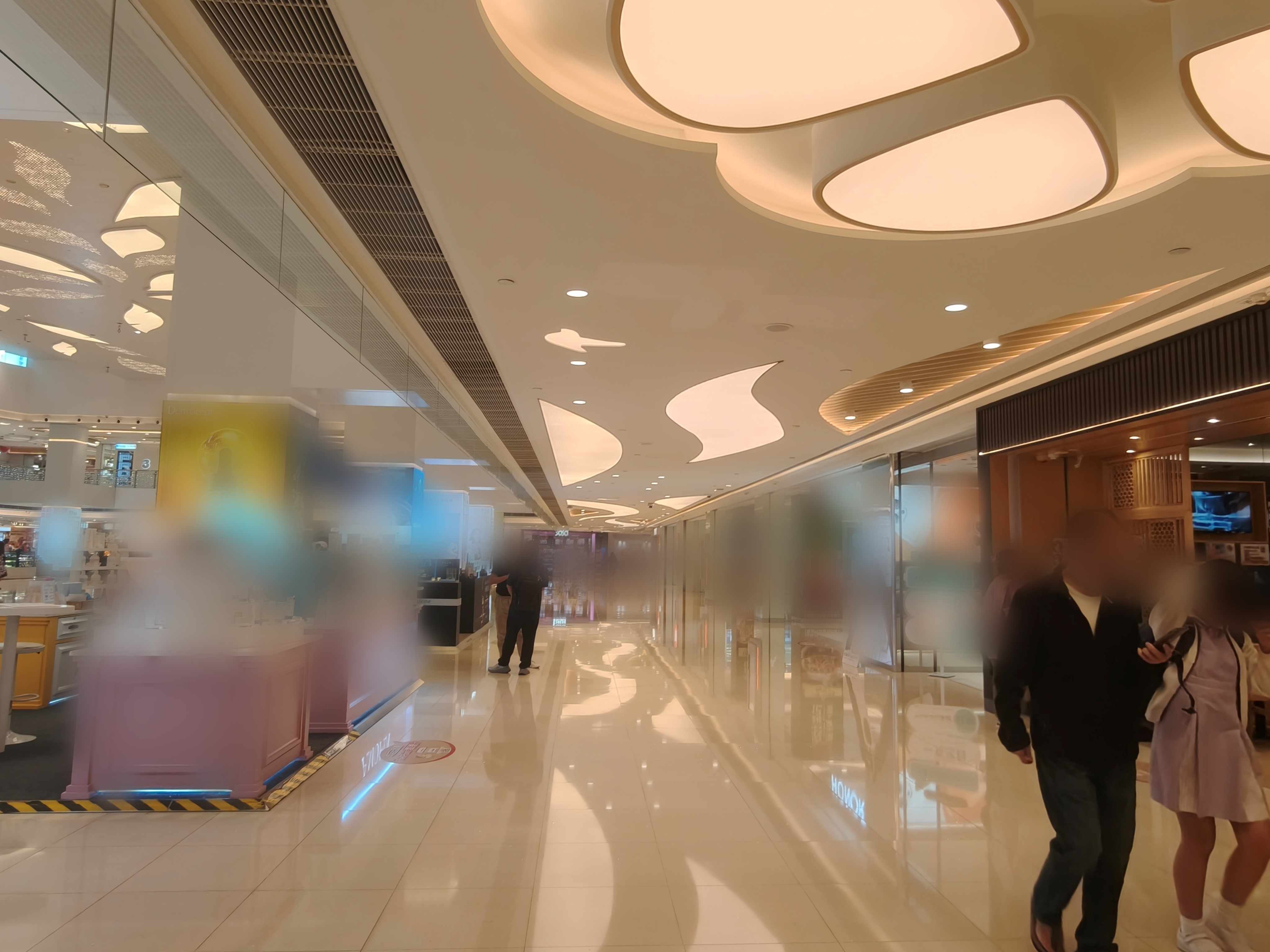}
        \caption{\blue{Mall}}
        \label{fig:mall}
    \end{subfigure}
    \hfill
    \begin{subfigure}[t]{0.24\linewidth}
        \centering
        \includegraphics[width=\linewidth]{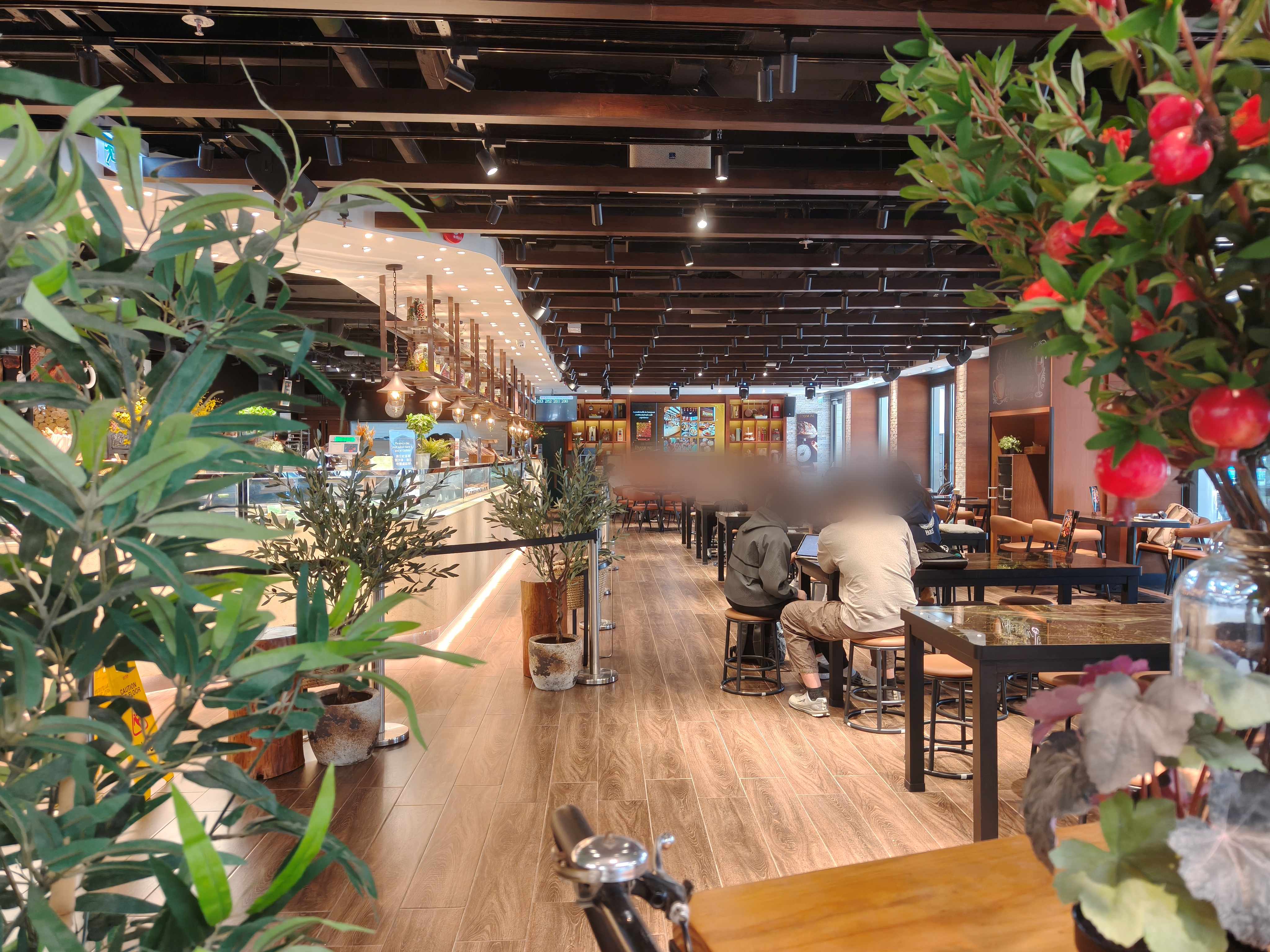}
        \caption{\blue{Cafeteria}}
        \label{fig:cafeteria}
    \end{subfigure}

    \vspace{2mm}

    \begin{subfigure}[t]{0.48\linewidth}
        \centering
        \includegraphics[width=\linewidth]{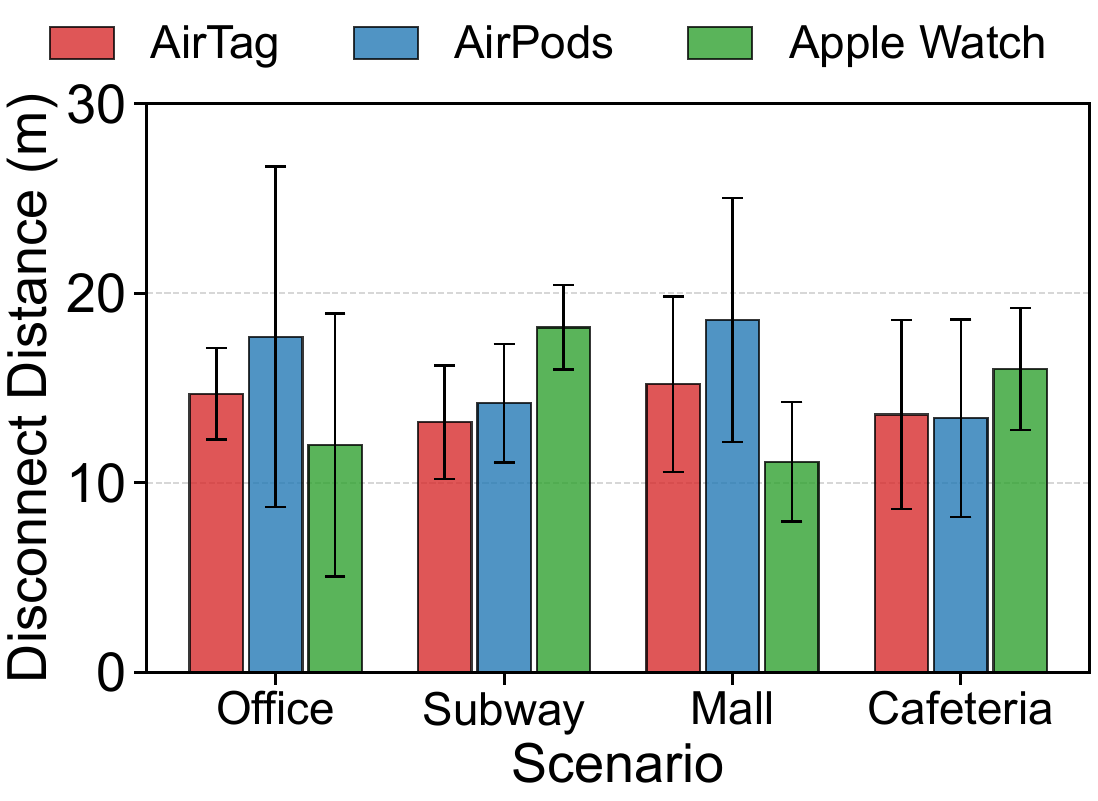}
        \caption{\blue{BLE disconnect distance (m) across different scenarios}}
        \label{fig:disconnect}
    \end{subfigure}
    \hfill
    \begin{subfigure}[t]{0.48\linewidth}
        \centering
        \includegraphics[width=\linewidth]{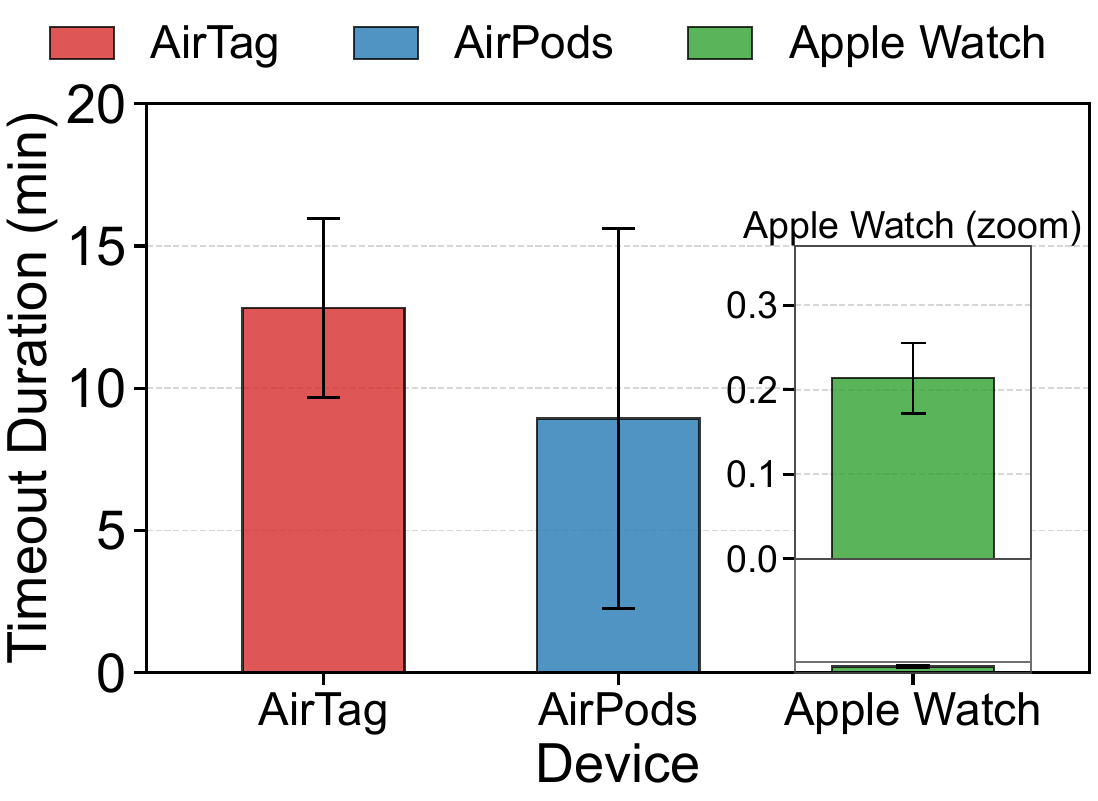}
        \caption{\blue{Timeout duration (min) for different devices}}
        \label{fig:timeout}
    \end{subfigure}
    \vspace{-2mm}
    \caption{\blue{Illustration of four scenarios for BLE disconnect distance evaluation (top), and the measured BLE disconnect distance and timeout duration for different devices to enter the \texttt{Separated (Lost)} state (bottom).}}
    \label{fig:scenarios_and_measurements}
    \vspace{-5mm}
\end{figure}

\noindent \textbf{\blue{Measurement of Essential State Parameters.}}
\blue{We measured two parameters governing the transition into the \texttt{Separated} state: the BLE disconnect distance and the timeout duration. The measurements were conducted across four representative environments: office, subway, mall, and cafeteria, as illustrated in \Cref{fig:office,fig:subway,fig:mall,fig:cafeteria}.}
\blue{For the BLE disconnect distance, as shown in \Cref{fig:disconnect}, it consistently falls within $11$--$19$ meters across all four scenarios (60 trials, 4 scenarios $\times$ 3 devices $\times$ 5 runs each).}
\blue{As for the timeout duration, it varies by device category, as shown in \Cref{fig:timeout}. AirTag averages $12.82 \pm 3.16$ minutes, AirPods $8.93 \pm 6.68$ minutes, and Apple Watch $0.21 \pm 0.04$ minutes (30 trials, 3 devices $\times$ 10 runs each).
These results indicate that Find My devices enter the \texttt{Separated} state within short BLE disconnect distances and timeout durations.
}

\subsection{Find My BLE Frame Structure}
\label{subsec:ble_frame}
\begin{figure}[t]
    \centering
    \includegraphics[width=\linewidth]{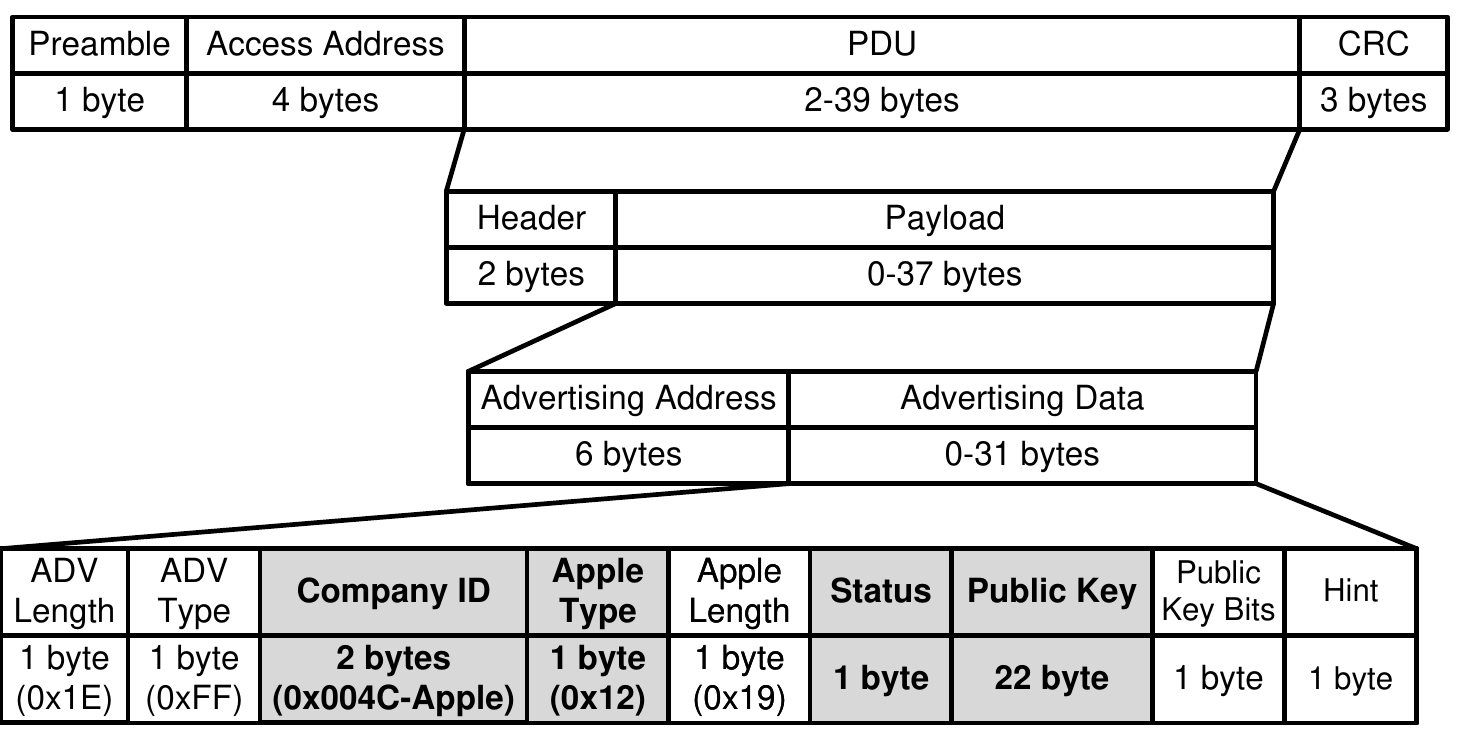}
    \caption{The frame structure of Find My BLE advertising frame triggered in separated mode. A full public key payload together Find My (0x12) Apple Type indicates the device is in the \texttt{Separated (Lost)} state.}
    \label{fig:findmy_bleframe}
\end{figure}

The Find My network relies on BLE advertisement frames to function \cite{heinrich2021can}.
As illustrated in Figure~\ref{fig:findmy_bleframe}, when a device enters the \texttt{Separated (Lost)} state, it continuously advertises these packets to announce its presence.

The structure adheres to standard BLE protocols but encodes proprietary logic within the \textit{Manufacturer Specific Data} field (set \textit{ADV Type} as 0xFF) to facilitate discovery:

\begin{itemize}
    \item \textbf{Company ID (0x004C):} Identifies the device manufacturer as Apple.
    \item \textbf{Find My Type (0x12):} Explicitly flags the advertises as a ``Find My'' packet. When accompanied by a complete public key payload, this indicates the device is in the \texttt{Separated (Lost)} state.
    \item \textbf{Status Byte:} Contains fine-grained device information, allowing receivers to distinguish specific device categories in the \texttt{Separated} state (e.g., differentiating an AirTag from an iPhone) \cite{10.1145/3507657.3528546}.
\end{itemize}

To protect user privacy, Apple devices employ a MAC address randomization scheme, periodically rotating the \textit{Advertising Address} to prevent long-term tracking. The payload also contains a rotating public key to ensure that only the owner can decrypt the location reports uploaded by Finders.

\subsection{Non-owner Sound Play Mechanism} 
\label{subsec:non_owner_sound}

A distinct feature of the Find My network is the permission for non-owners to interact with specific lost devices, particularly accessories. Apple explicitly allows surrounding devices to connect to a separated accessory (e.g., an AirTag) and trigger it to emit a sound.

This mechanism is designed primarily as an \textit{Anti-Stalking} feature. It enables potential victims to locate unknown trackers that may be moving with them without their knowledge. Technically, this interaction follows a standard Generic Attribute Profile (GATT) connection flow \cite{AppleFindMySpec2020}. A surrounding device can establish a BLE connection, perform service discovery, and write a specific command to a characteristic to trigger the alert, thereby revealing the physical location of the hidden accessory. 

\section{Threat Model}
\label{sec:threat_model}

In this section, we formally define the adversary model and analyze the specific security violations within the Find My network that facilitate the \textit{Snatcher} attack.

\subsection{Adversary Model}
We consider a physically proximate adversary whose objective is to \textbf{identify and navigate} to the lost Apple devices for theft. We assume a weak adversary model to demonstrate the severity of the vulnerability. Specifically, the adversary possesses the following characteristics:

\begin{itemize}
    \item \textbf{Equipment:} The attacker utilizes commodity hardware (COTS), such as a standard Android smartphone. No specialized equipment (e.g., Software Defined Radios or directional antenna arrays) is required.
    \item \textbf{Knowledge:} The attacker has zero prior knowledge of the target. They possess no cryptographic keys, are not part of the owner's ``Family Sharing'' group, and have no access to the owner's iCloud account.
    \item \textbf{Capability:} The attacker can passively scan BLE advertisements and actively initiate standard GATT connections. 
\end{itemize}

\blue{The attacker aims to leverage the digital presence of a lost device (e.g., BLE signals or acoustic feedback) to navigate towards the target, bridging the gap between digital detection and physical acquisition.} \blue{This adversary model is practical because everyday scenarios such as leaving a bag behind, or walking away from a table frequently create the conditions for attack. As established in Section~\ref{subsec:operation_flow}, devices enter the vulnerable \texttt{Separated (Lost)} state within $15$ minutes and at short separation distances ($11$--$19$ m), making these everyday separations sufficient for exploitation.}

\subsection{Vulnerability Analysis}
Based on the operational logic described in Section~\ref{subsec:operation_flow}, we identify three fundamental vulnerabilities where the system's design goals (Usability and \textit{Anti-Stalking}) conflict with physical security (\textit{Anti-Theft}).

\subsubsection{Vulnerability I: Cleartext Discovery Beacon (Failure of Obfuscation)}
\label{subsec:vuln1}
While the content of location reports is encrypted, the \textit{existence} of a lost device is advertising in cleartext. The Find My protocol uses a static identifier (Type \texttt{0x12}) in the BLE Manufacturer Specific Data to ensure universal discoverability.

\textbf{Implication:} This design violates the principle of obfuscation. It allows an adversary to implement a simple pre-filtering mechanism, instantly distinguishing high-value \texttt{Seperated (Lost)} targets from thousands of background Bluetooth devices without needing to attempt a connection. This eliminates the need for visual search, allowing attackers to stealthily identify potential targets.

\subsubsection{Vulnerability II: Unauthenticated Actuation (The Acoustic Side-Channel)}
\label{subsec:vuln2}
To mitigate stalking risks, Apple allows unauthenticated devices to trigger a sound on \texttt{Separated} accessories (e.g., AirTags).

\textbf{Implication:} This creates a physically perceptible acoustic side-channel.
Crucially, our investigation reveals that the capability to actuate this sound is device-agnostic.
As shown in Table~\ref{tab:non_owner_sound_play}, standard GATT commands can trigger this sound from generic commodity hardware, including Android phones and ESP32 boards.
This effectively weaponizes the intended \textit{Anti-Stalking} safeguard to facilitate theft.
The emitted sound provides the attacker with immediate auditory confirmation of the device's presence and enables precise acoustic localization to pinpoint the hidden item (e.g., inside a specific pocket of a bag).

\begin{table}[h]
\centering
\caption{Summary of the capability of triggering sound on lost Apple devices using various attacker devices.}
\vspace{-2mm}
\label{tab:non_owner_sound_play}
\setlength{\tabcolsep}{3pt}
\begin{tabular}{lcccc}
\toprule
\textbf{Non-owner} & \textbf{AirTag} & \textbf{AirPods} & \textbf{iPhone} & \textbf{Apple} \\
\textbf{Device} & & & & \textbf{Watch} \\
\midrule
Google Pixel 7 & $\checkmark$ & $\checkmark$ & $\times$ & $\times$ \\
OPPO Reno 10 & $\checkmark$ & $\checkmark$ & $\times$ & $\times$ \\
Huawei Nova 12s & $\checkmark$ & $\checkmark$ & $\times$ & $\times$ \\
ESP32-WROVER-E & $\checkmark$ & $\checkmark$ & $\times$ & $\times$ \\
\bottomrule
\multicolumn{5}{l}{\scriptsize $\checkmark$: Successfully triggered sound; $\times$: Connection refused or command ignored.}
\end{tabular}
\end{table}

\subsubsection{Vulnerability III: Extended Identity Persistence}
\label{subsec:vuln3}
While MAC address randomization is employed to prevent long-term tracking, the update period is excessively long.

\textbf{Implication:} As detailed in Table~\ref{tab:mac_shuffle_period}, 
the address rotation period (e.g., approximately 24 hours for AirTags) far exceeds the duration of a typical theft operation. This prolonged static period is likely an intentional design choice to facilitate \textit{Anti-Stalking} detection, which requires observing a consistent identity over time to identify unwanted tracking. However, for a thief, this creates a stable window of opportunity. The persistence of the ID allows the attacker to continuously observe signal gradients and navigate towards the signal peak without the risk of the target disappearing due to an address change.

\begin{table}[h]
\centering
\caption{Measured MAC address rotation periods for different Apple devices in Separated (Lost) state.}
\label{tab:mac_shuffle_period}
\begin{tabular}{llc}
\toprule
\textbf{Category} & \textbf{Device} & \textbf{MAC Address Rotation Period} \\
\midrule
\multirow{2}{*}{\begin{tabular}[c]{@{}l@{}}Apple\\ Accessories\end{tabular}} 
 & AirPods Pro & $\sim$24 hours (updated at 4 am) \\
 & AirTag      & $\sim$24 hours (updated at 4 am)\\ 
\midrule
\multirow{2}{*}{\begin{tabular}[c]{@{}l@{}}Apple\\ Devices\end{tabular}}     
 & iPhone      & Fixed 15 mins \\
 & Apple Watch & 19--36 mins \\ 
\bottomrule
\end{tabular}
\end{table}

\subsection{The Snatcher Attack Framework}
Exploiting these vulnerabilities, we propose \textit{Snatcher}, a progressive three-level attack framework tailored to different device capabilities and defense levels.

\begin{itemize}
    \item \textit{Level 1: Acoustic Direction Finding.} Targeting accessories (e.g., AirTags), this Level exploits Vulnerability II. The attacker forces the device to emit sound and utilizes a \textit{Acoustic Direction Finding} algorithm. This method provides an explicit physical vector, allowing the attacker to approach the target efficiently even in non-line-of-sight conditions.
    
    \item \textit{Level 2: RSSI-IMU Navigation via Sensor Fusion.} Targeting silent devices that resist sound triggers (e.g., iphone), this stage exploits Vulnerability III. Since the target is silent, the attacker relies solely on RF signal trends. We propose an \textit{RSSI-IMU Navigation} algorithm that fuses noisy RSSI variations with the attacker's walking trajectory (IMU data). By analyzing the correlation between movement and signal strength changes, the system provides step-by-step guidance.
    
    \item \textit{Level 3: De-anonymization via Spatial-Temporal Clustering.} Addressing advanced defenses where MAC addresses might rotate rapidly (emulated scenario), this stage challenges the assumption that identity rotation prevents physical tracking. We propose a \textit{Spatial-Temporal Clustering} strategy that stiches discontinuous signal segments. This allows the navigation process to continue seamlessly even when the target's digital identity changes mid-attack.
\end{itemize}

\section{Design}
\label{sec:design}

This section details the design of \textit{Snatcher}, a progressive three-level attack model. It begins with \textit{Level 1: Acoustic Direction Finding} (Section~\ref{sec:stage1_design}), which uses sound to locate vulnerable accessories. Next, \textit{Level 2: RSSI-IMU Navigation} (Section~\ref{sec:stage2_design}) employs an uncertainty-aware Bayesian fusion of signal strength and motion data to navigate silent devices. Finally, \textit{Level 3: De-anonymization via Spatial-Temporal Clustering} (Section~\ref{sec:stage3_design}) defeats MAC address randomization by re-associating device identities.

\subsection{Level 1: Acoustic Direction Finding}
\label{sec:stage1_design}

\begin{figure}[t]
    \centering
    \includegraphics[width=\linewidth]{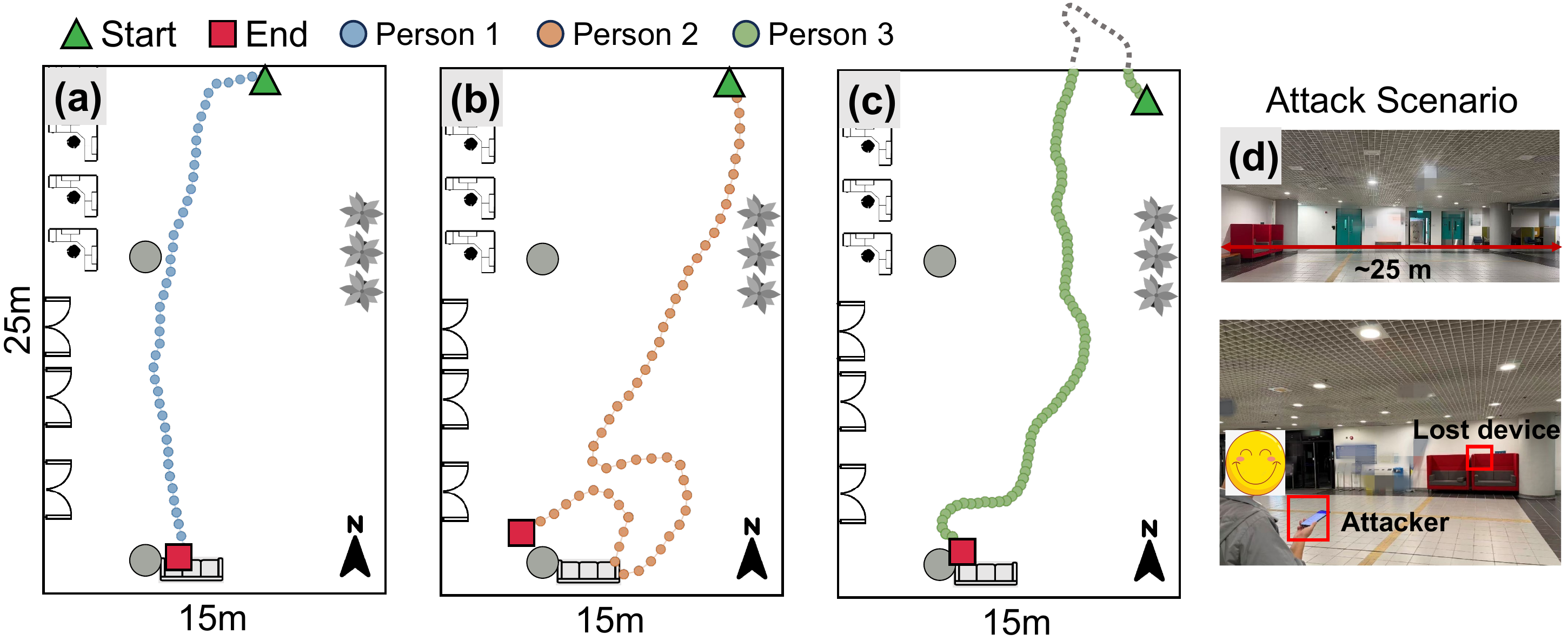}
    \caption{Examples demonstrating preliminary validation of a user's ability to hear and navigate to a lost device using the non-owner sound playback feature. (a) Direct finding. (b) \& (c) Cases where attackers cannot determine the correct bearing to the sound source and conduct multiple random search attempts. (d) Attack scenario in an indoor office-like open area.}
    \label{fig:user_study}
\end{figure}

The objective of Level 1 is to guide the attacker from a state of total location uncertainty to the immediate vicinity of the lost accessory. 
Relying solely on human hearing for this task is often inefficient, since auditory localization based on binaural and spectral cues (e.g., interaural time and level differences) is highly sensitive to the acoustic context and individual variability \cite{KEATING201535,Romigh2014Individualized}.

To demonstrate this limitation in our setting, we conducted preliminary tests in a representative indoor environment—a student office-like open area with a mix of Line-of-Sight (LoS) pathways and numerous sources of Non-Line-of-Sight (NLoS) interference, such as furniture and pedestrian occlusions—as shown in Figure~\ref{fig:user_study}d. While direct navigation in such an environment is possible (Figure~\ref{fig:user_study}a), we observed frequent and significant errors rooted in the misjudgment of the initial direction. For instance, even a slight initial heading error can cause the user to become disoriented, leading to hesitant, looping movements before finding the correct path (Figure~\ref{fig:user_study}b). A more severe failure occurs when the user initially chooses a completely opposite direction, resulting in a significantly prolonged and inefficient search trajectory (Figure~\ref{fig:user_study}c).

This inherent unreliability motivates an \textit{acoustic direction finding} system that offers a consistent and user-independent way to obtain a reliable initial heading, replacing human guesswork with algorithmic guidance. The process is divided into two phases: \textit{Initial Direction Estimation} and \textit{Step-wise Navigation}.

\begin{figure*}[t]
    \centering
    \begin{subfigure}[b]{0.32\textwidth}
        \centering
        \includegraphics[width=.8\textwidth]{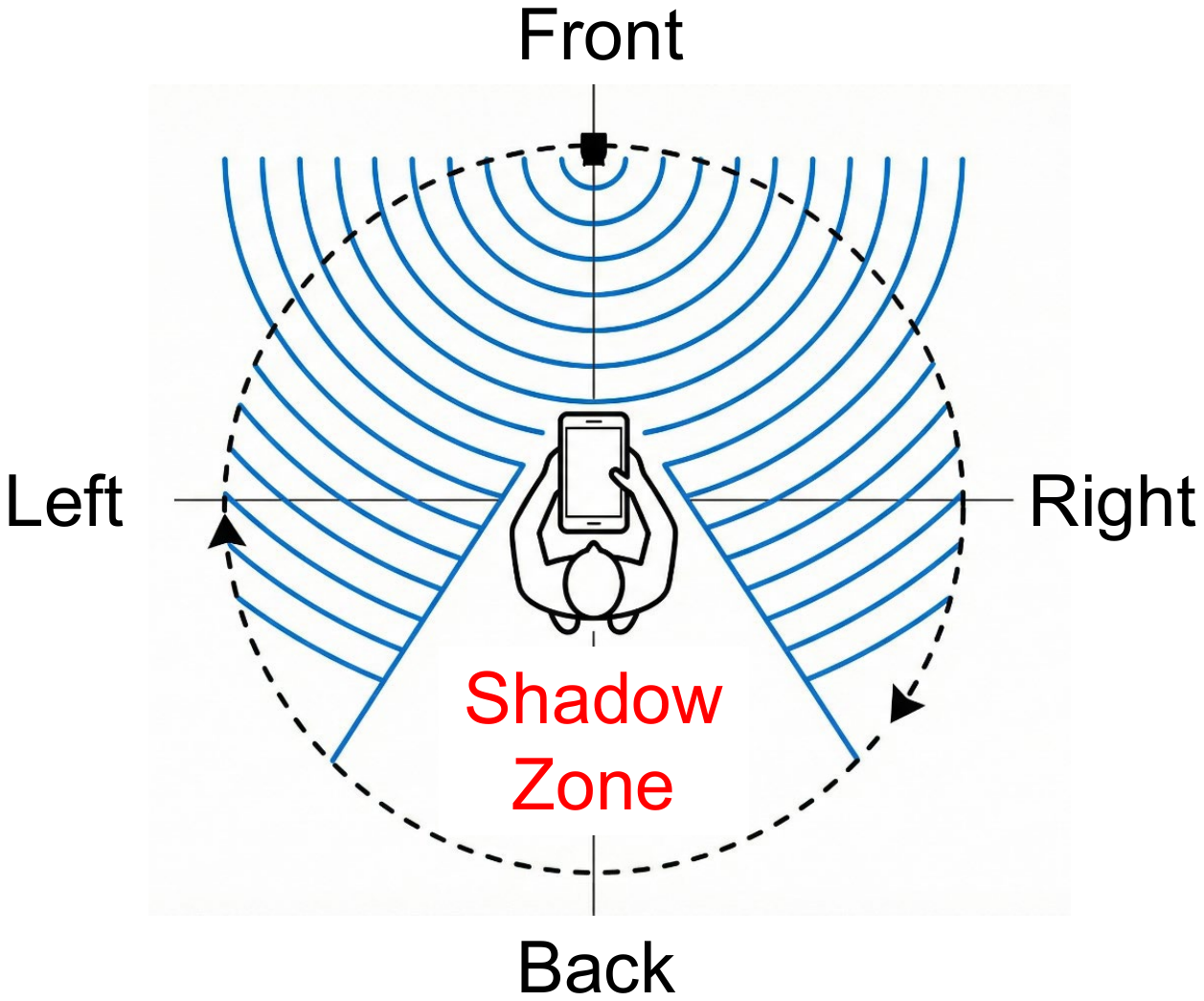} 
        \caption{Human Body Shadowing Principle}
        \label{fig:stage1_concept}
    \end{subfigure}
    \hfill
    \begin{subfigure}[b]{0.3\textwidth}
        \centering
        \includegraphics[width=.8\textwidth]{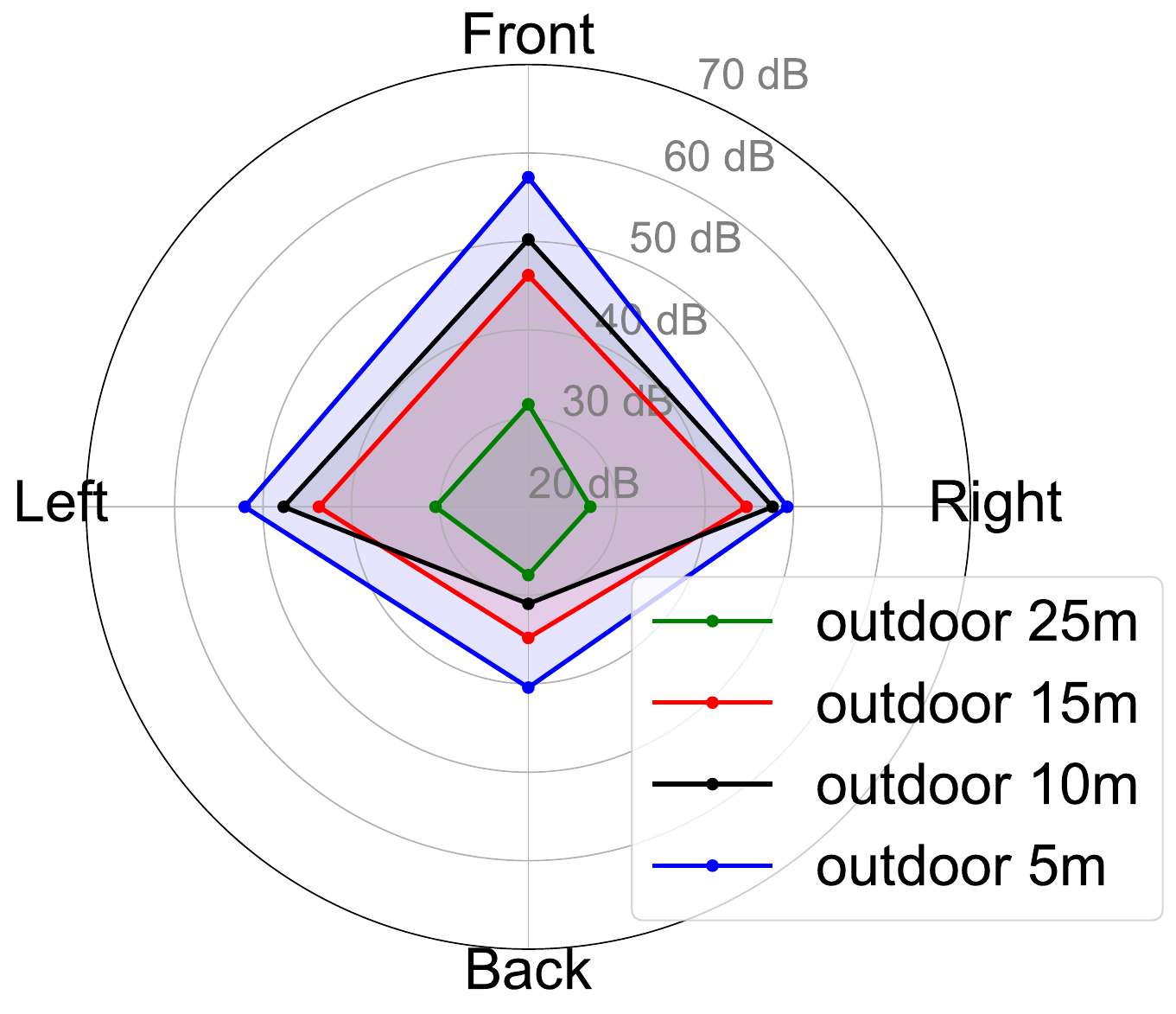}
        \caption{Outdoor Measurements}
        \label{fig:stage1_outdoor}
    \end{subfigure}
    \hfill
    \begin{subfigure}[b]{0.3\textwidth}
        \centering
        \includegraphics[width=.8\textwidth]{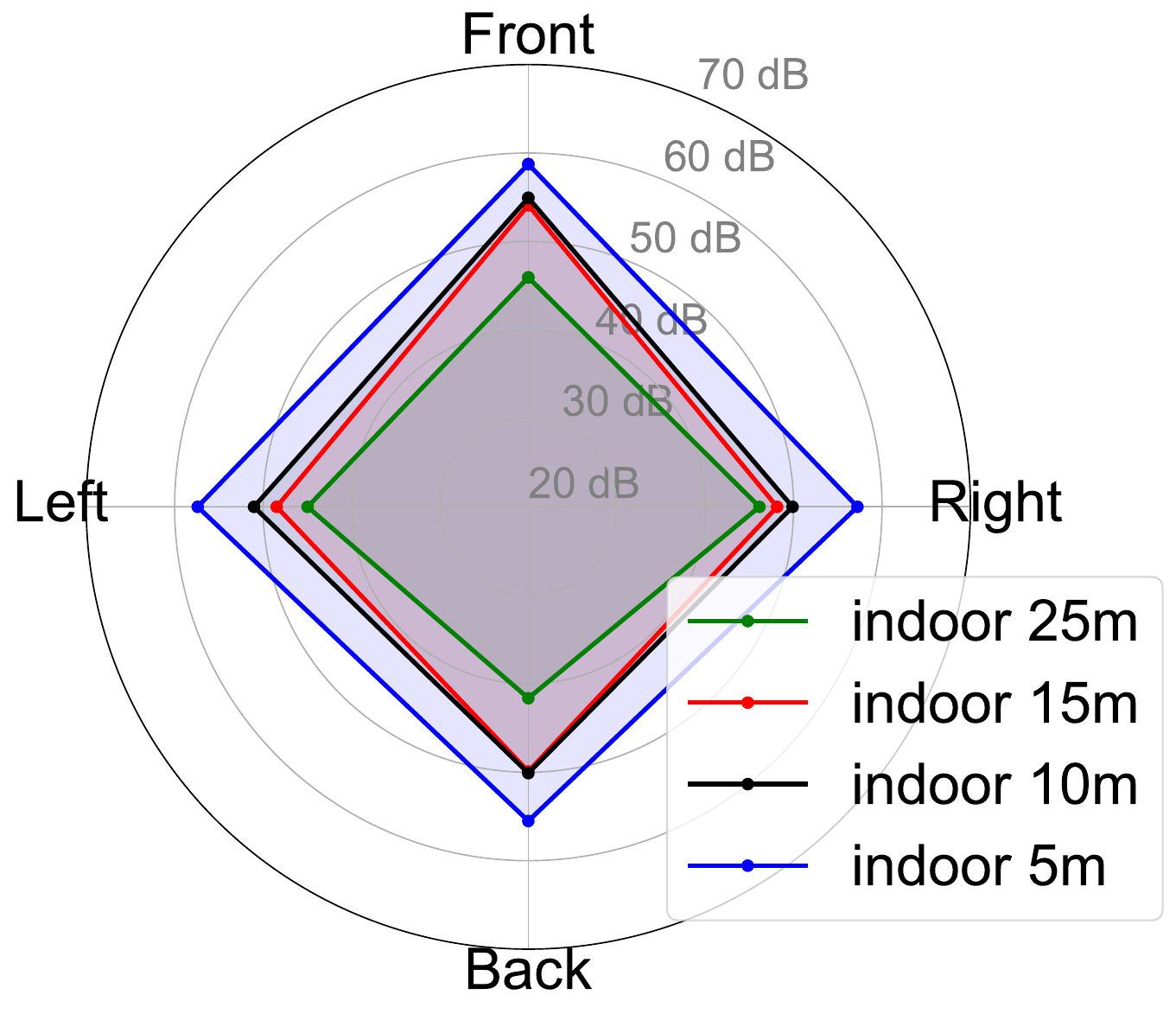}
        \caption{Indoor Measurements} 
        \label{fig:stage1_indoor}
    \end{subfigure}
    
    \caption{\textbf{Physical intuition and verification of human body shadowing.} (a) A top-down schematic illustrating the acoustic shadow zone created behind the user. (b) \& (c) Measured signal strength (dB) polar plots in outdoor and indoor environments. The distinct signal drop at Back validates the feasibility of using body shadowing for direction finding.}
    \label{fig:stage1_combined}
    \vspace{-1mm}
\end{figure*}

\subsubsection{Phase I: Initial Direction Estimation}

Before moving, the attacker first establish a coarse heading vector towards the sound source. In the absence of a directional microphone array, we propose a method that leverages the attacker's own body as a physical shield to create a virtual directional antenna, which forms the basis of our initial direction estimation procedure.

\noindent
\textbf{Physical Intuition (Human Body Shadowing).}
The acoustic signal emitted by Apple accessories (e.g., AirTags) typically consists of high-frequency chirps in the $2\text{kHz} \sim 4\text{kHz}$ range. At these frequencies, the wavelength ($\lambda \approx 8.5 \sim 17 \text{cm}$) is significantly smaller than the width of an adult human torso ($D > 30 \text{cm}$). According to diffraction theory, when $D > \lambda$, the obstacle creates a distinct \textit{Shadow Zone} behind it, as illustrated in Figure~\ref{fig:stage1_concept}.
Consequently, when the attacker holds the smartphone against their chest, the body effectively blocks direct signal paths from the rear, creating a cardioid-like sensitivity pattern.

This phenomenon is validated by our measurements in real-world scenarios. Figure~\ref{fig:stage1_outdoor} and Figure~\ref{fig:stage1_indoor} present the measured signal strength patterns in outdoor and indoor environments, respectively.
Observations indicate that: 
1) In the outdoor scenario (Figure~\ref{fig:stage1_outdoor}), the signal exhibits a sharp attenuation of approximately $20\sim30$ dB at the back orientation compared to the Front.
2) Crucially, in the indoor scenario (Figure~\ref{fig:stage1_indoor}), despite multipath reflections and reverberations, the body shadow effect remains robust. A clear signal dip (approx. $10\sim15$ dB) is consistently observed at the Back direction.
This persistence of the Front-Back differential, even in complex indoor environments, holds as a reliable physical basis of our estimation.

\noindent
\textbf{Measurement Operation.} To exploit this phenomenon, we implement a \textit{Scan \& Spin} heuristic. The attacker holds the smartphone tightly against their chest to utilize the body shield and rotates through four orthogonal orientations ($0^\circ, 90^\circ, 180^\circ, 270^\circ$). At each orientation $k \in \{F, R, B, L\}$, the attacker triggers the sound and records the received signal strength in Decibels (dB), denoted as $A_k$.

\noindent
\textbf{Contrast-Enhanced Directionality Metric.}
We further design a direction finding metric by leveraging
two key physical insights: (1) the correct direction should exhibit high signal intensity; and (2) due to body shadowing, the correct direction should maximize the signal drop when the user turns $180^\circ$ away ($A_k - A_{opp(k)}$). Based on these principles, we formulate the \textit{Contrast-Enhanced Score} ($S_k$) as a weighted linear combination:
\begin{equation}
    S_k = \alpha \cdot A_k + \beta \cdot (A_k - A_{opp(k)})
\end{equation}
Here, $A_{opp(k)}$ denotes the signal strength (dB) measured at the opposite orientation (e.g., Back for Front). The parameters $\alpha$ and $\beta$ serve as weighting coefficients for the absolute intensity and the directional contrast, respectively. We constrain the weights such that $\alpha + \beta = 1$ (experimentally set to $\alpha=0.3, \beta=0.7$ to prioritize the directional differentiation). The estimated initial direction is then determined by $D_{init} = \arg\max_k (S_k)$.

\begin{figure}[t]
    \centering
    \includegraphics[width=\linewidth]{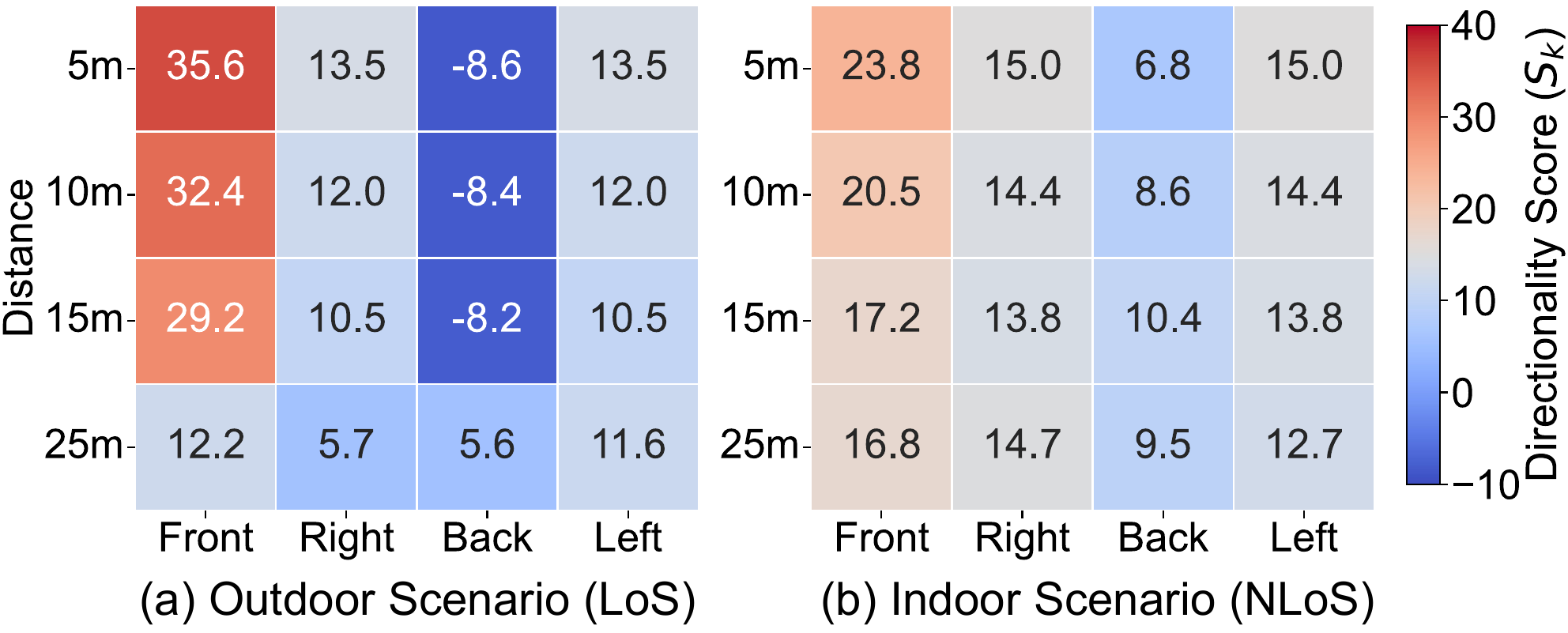}
    
\caption{\textbf{Heatmap of Directionality Scores ($S_k$).} 
The \textit{Front} direction consistently achieves the highest score across all scenarios. This holds true even in the challenging indoor environment, validating the robustness of our metric.}
    \label{fig:direction_heatmaps}
    \vspace{-2mm}
\end{figure}

\noindent
\textbf{Metric Verification.}
We apply the optimized metric to the outdoor and indoor measurement~(statistics in Figure~\ref{fig:stage1_combined}). The heatmap in Figure~\ref{fig:direction_heatmaps} visualizes the results. 

As shown in Figure~\ref{fig:direction_heatmaps}a, the heatmap reveals a distinct hot zone along the Front direction. The high contrast between the Front (red) and Back (blue) regions confirms that the body shadow successfully creates a sharp directional gradient in ideal conditions. In the complex indoor environment (Figure~\ref{fig:direction_heatmaps}b), multipath reflections tend to blur signal boundaries. However, the metric amplifies the subtle Front-Back difference. Consequently, the Front direction remains the discernible global maximum (highest heat intensity) across all distances, effectively suppressing the ambiguous side directions.

\subsubsection{Phase II: Step-wise Navigation}

Once the initial heading $D_{init}$ is established via the directionality metric ($S_k$), the attacker enters the navigation phase. Given the active nature of the acoustic trigger, we employ a \textit{Move-Verify-Correct} strategy to efficiently guide the attacker while minimizing interaction overhead.

The process operates as an iterative feedback loop. The attacker advances along the estimated heading for a discrete distance and then halts to trigger a single sound feedback. 
The system compares the new signal amplitude ($A_{new}$) against the previous reading ($A_{prev}$). If the signal strengthens ($A_{new} > A_{prev} + \delta$), confirming the correct path, the system instructs the attacker to maintain the current heading. Conversely, if the signal significantly drops, indicating a deviation or overshoot, the system triggers a state correction. In this case, the attacker is prompted to halt and re-execute the \textit{Phase I} \textit{Scan \& Spin} procedure. By recalculating the full contrast-enhanced metric $S_k$, the system re-calibrates the heading vector towards the true source before resuming movement. This hybrid approach prioritizes speed by relying on simple amplitude checks during linear approach, while reserving the robust (but slower) $S_k$ metric for necessary course corrections.

\begin{figure}[t]
    \centering
    \includegraphics[width=\linewidth]{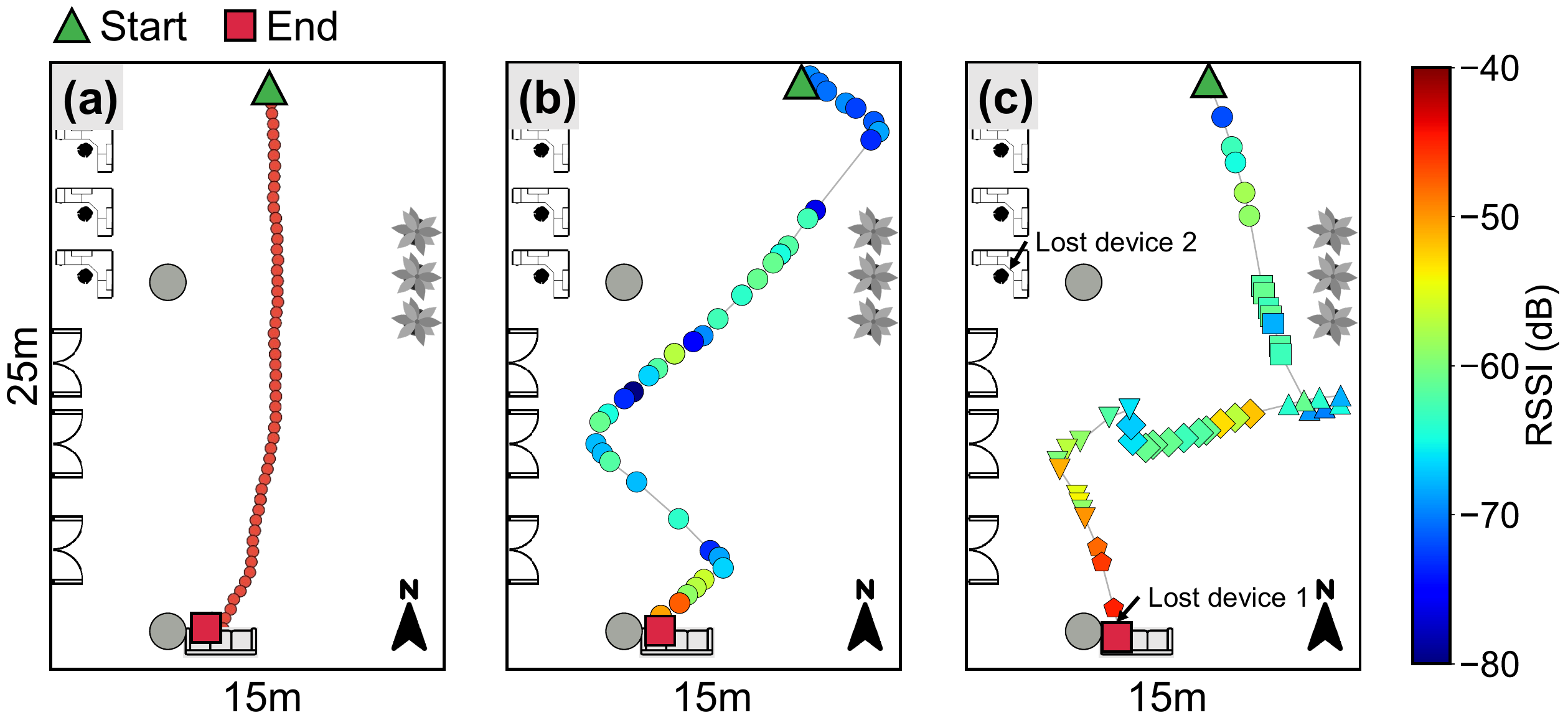}
    \caption{Experimental validation of multi-level navigation.
(a) Acoustic direction-finding-based navigation.
(b) RSSI-IMU navigation.
(c) RSSI-IMU navigation with spatial-temporal clustering, involving two lost devices that rotate their MAC addresses every 30 seconds.
In (b) \& (c), scatter points with different colors represent RSSI strength, while points with different shapes correspond to distinct MAC addresses.}
    \label{fig:all_three}
\vspace{-6mm}
\end{figure}

\textbf{Preliminary Verification.} Figure~\ref{fig:all_three}a illustrates a representative navigation trial in the indoor environment shown in Figure~\ref{fig:user_study}d, where our acoustic guidance yields an accurate initial heading and a visibly more direct path to the target than unassisted human hearing (Figure~\ref{fig:user_study}b,c).
This behavior primarily stems from the \textit{Scan \& Spin} procedure, which effectively acquires a reliable initial heading for navigation, while the amplitude-based \textit{Move-Verify-Correct} strategy then helps the user maintain the correct direction and promptly revise erroneous headings during the walk.

\subsection{Level 2: RSSI-IMU Navigation via Uncertainty-Aware Bayesian Fusion}
\label{sec:stage2_design}

For silent devices (e.g., iPhones) that do not respond to acoustic triggers, the attacker’s Android phone adopts a purely passive navigation strategy that observes only the victim’s BLE advertisements and the phone’s own inertial sensors. In this setting, the external information is limited to the RSSI values of the lost device’s broadcasts, while the attacker’s IMU is used to reconstruct the walking trajectory and heading over time.
A fundamental challenge in such RSSI-based navigation is the trade-off between \textit{sensitivity} and \textit{robustness}. Relying on instantaneous RSSI changes (\textit{short-term}) makes the system highly responsive to local direction changes, but also extremely vulnerable to multipath fading and random signal spikes. In contrast, relying on cumulative averages (\textit{long-term}) effectively smooths out noise, yet introduces substantial lag, causing the navigation to react too slowly to sudden U-turns or course corrections.

\begin{figure}[t]
    \centering
    \includegraphics[width=.98\linewidth]{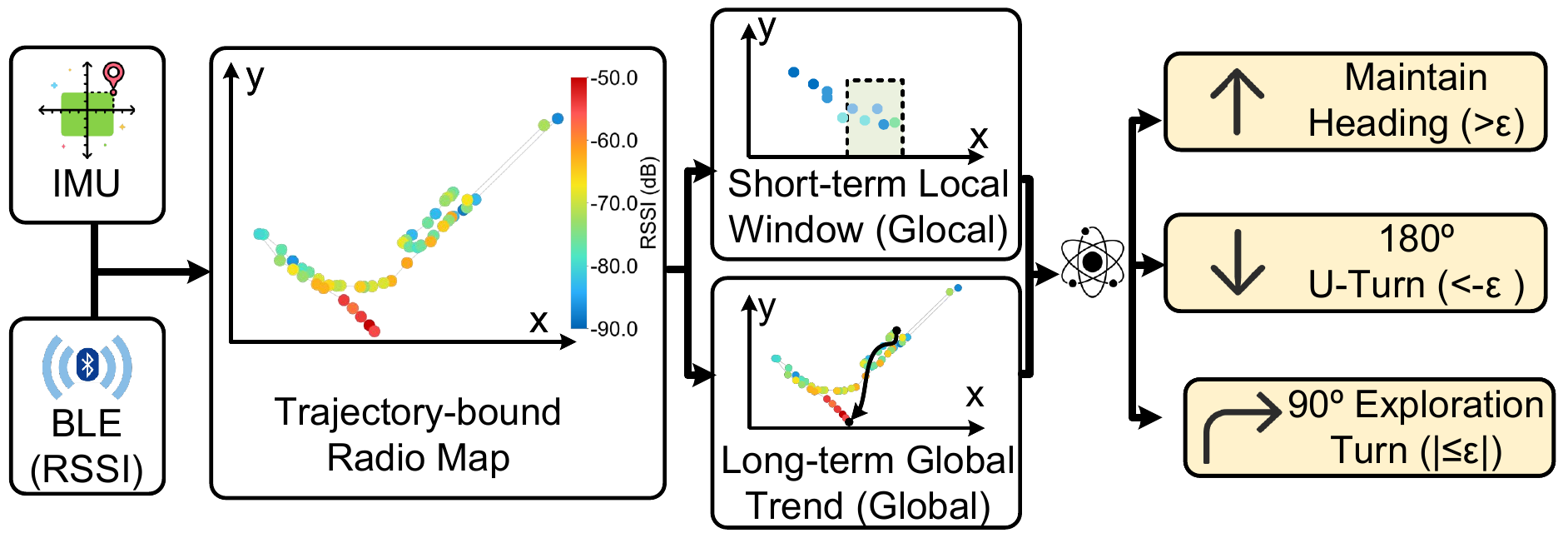}
    \caption{Architecture of the RSSI-IMU Navigation System. (Left) IMU and BLE inputs construct a trajectory-bound radio map. (Center) Dual-scale processing extracts the Short-term Local Gradient and Long-term Global Trend. (Right) An uncertainty-aware Bayesian node fuses these estimates for robust guidance.}
    \label{fig:stage2}
\end{figure}

To resolve this dichotomy, we propose an \textit{Uncertainty-Aware Bayesian Fusion} system. As illustrated in Figure~\ref{fig:stage2}, our approach reconstructs a spatial trajectory and dynamically fuses signal trends from the \textit{long-term} and \textit{short-term} temporal scales based on their statistical confidence.

\subsubsection{Spatial-Signal Trajectory Mapping}
\label{sec:mapping}
The first step is to convert the time-domain RSSI measurements to its corresponding spatial representation, as signal propagation is a function of physical distance.
We adopt a Pedestrian Dead Reckoning (PDR) \cite{rashid2015dead} engine that fuses accelerometer, gyroscope, and magnetometer data to reconstruct the user's relative spatial trajectory. 
By mapping the RSSI of each received BLE packet $r_k$ to its estimated physical coordinate $\mathbf{p}_k$, we construct a \textit{Trajectory-bound Radio Map}, denoted as $\mathcal{T}_k = \{(\mathbf{p}_1, r_1), \dots, (\mathbf{p}_k, r_k)\}$. This spatial alignment enables the extraction of gradients based on physical geometry.

\subsubsection{Dual-Scale Trend Analysis}
Based on the mapped trajectory $\mathcal{T}_k$, the system extracts navigation gradients from two complementary perspectives:

\textit{1. Short-term: Robust Local Gradient ($G_{local}$).}
This component mimics the \textit{exploration} behavior, focusing on immediate responsiveness. We employ a sliding window regression over the most recent trajectory segment (e.g., the last 3-5 meters). 
The output includes the local slope estimate $G_{local}$ and its variance $\sigma^2_{local}$. A high $\sigma^2_{local}$ indicates a chaotic local RF environment (e.g., deep fading), signaling low confidence in the immediate trend.

\textit{2. Long-term: Global Trend Fitting ($G_{global}$).}
This component mimics the \textit{experience} behavior, aiming to construct a macroscopic understanding of the signal propagation environment. As the attacker navigates, the accumulated sequence of spatial-signal pairs $\mathcal{T}_k$ effectively forms a representative radio map of the traversed area. While local measurements fluctuate due to multipath effects, the aggregate signal distribution over a long trajectory reveals the underlying global trend of the RSSI spatial distribution.

We exploit this by extracting a continuous trendline that effectively cuts through the scattered raw data points to identify the direction of the maximized spatial gradient, as illustrated in the bottom branch of Figure 8. By filtering out small-scale fading over long distances, this curve exposes the stable \textit{Global Gradient} ($G_{global}$), serving as a macro-scale anchor for the heading.
Specifically, we calculate $G_{global}$ as the spatial derivative of this smoothed RSSI distribution. A positive $G_{global}$ confirms that, despite potential local dips, the user's overall trajectory is converging towards the signal source.
Crucially, we model the uncertainty of this estimate, $\sigma^2_{global}$, to be inversely proportional to the trajectory length $L_k$. This mathematically captures the intuition that \textit{more data yields higher confidence}: as the radio map expands, the system's grasp of the global trend becomes increasingly robust against local anomalies.

\subsubsection{Uncertainty-Aware Bayesian Fusion}
To combine these two complementary indicators, a naive approach might use fixed weights (e.g., average) or heuristic time-decay functions. However, such methods fail to capture the \textit{quality} of the estimation at any given instant. Instead, we formulate the fusion as a Maximum Likelihood Estimation (MLE) problem under Gaussian uncertainty. We employ Inverse-Variance Weighting, a Bayesian method that optimally combines independent estimators by weighing them according to their precision:

\begin{equation}
    G_{final}(k) = \frac{\sigma^2_{global} \cdot G_{local} + \sigma^2_{local} \cdot G_{global}}{\sigma^2_{local} + \sigma^2_{global}}
\end{equation}

As we see, the contribution of each gradient component is inversely proportional to its uncertainty ($\sigma^2$), meaning that the estimator with higher precision (lower variance) mathematically dominates the final decision.

More importantly, this formulation naturally enables an \textit{evolutionary and adaptive weighting policy} without manual parameter tuning. First, it establishes \textit{Evolutionary Trust}: at the start of navigation ($k$ is small), the global history is sparse and unreliable ($\sigma^2_{global} \to \infty$). The formula automatically assigns dominance to $G_{local}$, ensuring the system remains responsive during the initial search phase. As data accumulates, the confidence in the global map increases ($\sigma^2_{global} \to 0$), and the system seamlessly shifts to rely on the robust global trend. Second, it provides inherent \textit{Noise Gating}: if the attacker steps into a signal null, the local regression variance $\sigma^2_{local}$ spikes due to the poor linear fit. The Bayesian formula immediately down-weights the local input, effectively locking the decision onto the stable global trend, thereby preventing false U-turn instructions caused by temporary signal fluctuations.

The final fused gradient $G_{final}$ drives the navigation system to provide guidance that is both agile in clean environments and robust in complex multipath conditions.

\subsubsection{Navigation Feedback Policy}
To translate the scalar gradient $G_{final}$ into actionable guidance, we employ a threshold-based logic with a noise deadband $\epsilon$. A positive gradient ($G_{final} > \epsilon$) confirms an approaching trend, guiding the attacker to maintain the current heading. Conversely, a significant negative gradient ($G_{final} < -\epsilon$) indicates deviation, prompting an immediate $180^{\circ}$ U-Turn correction. In cases where the gradient plateaus ($|G_{final}| \le \epsilon$), suggesting tangential movement or a signal null, the system advises an exploratory $90^{\circ}$ turn to re-acquire a distinct trend. 
After collecting sufficient RSSI samples, the exploratory turn is directed towards the direction of the maximum global gradient, enabling faster convergence to the correct path.
This logic establishes a continuous feedback loop, enabling the attacker to perform a physical gradient ascent towards the target.

\noindent
\textbf{Preliminary Verification.} To demonstrate the efficacy of this system, we conducted a navigation task against a silent device in the same challenging indoor environment (Figure~\ref{fig:user_study}d). As shown in Figure~\ref{fig:all_three}b, the system successfully guided the user to the target. The trajectory color gradient, transitioning from blue (low RSSI) to red (high RSSI), visualizes a clear spatial progression. This result confirms that our Uncertainty-Aware Bayesian Fusion can effectively filter transient fluctuations and lock onto the true spatial gradient, enabling robust navigation even in a complex multipath environment.

\begin{figure}[t]
    \centering
    \includegraphics[width=.99\linewidth]{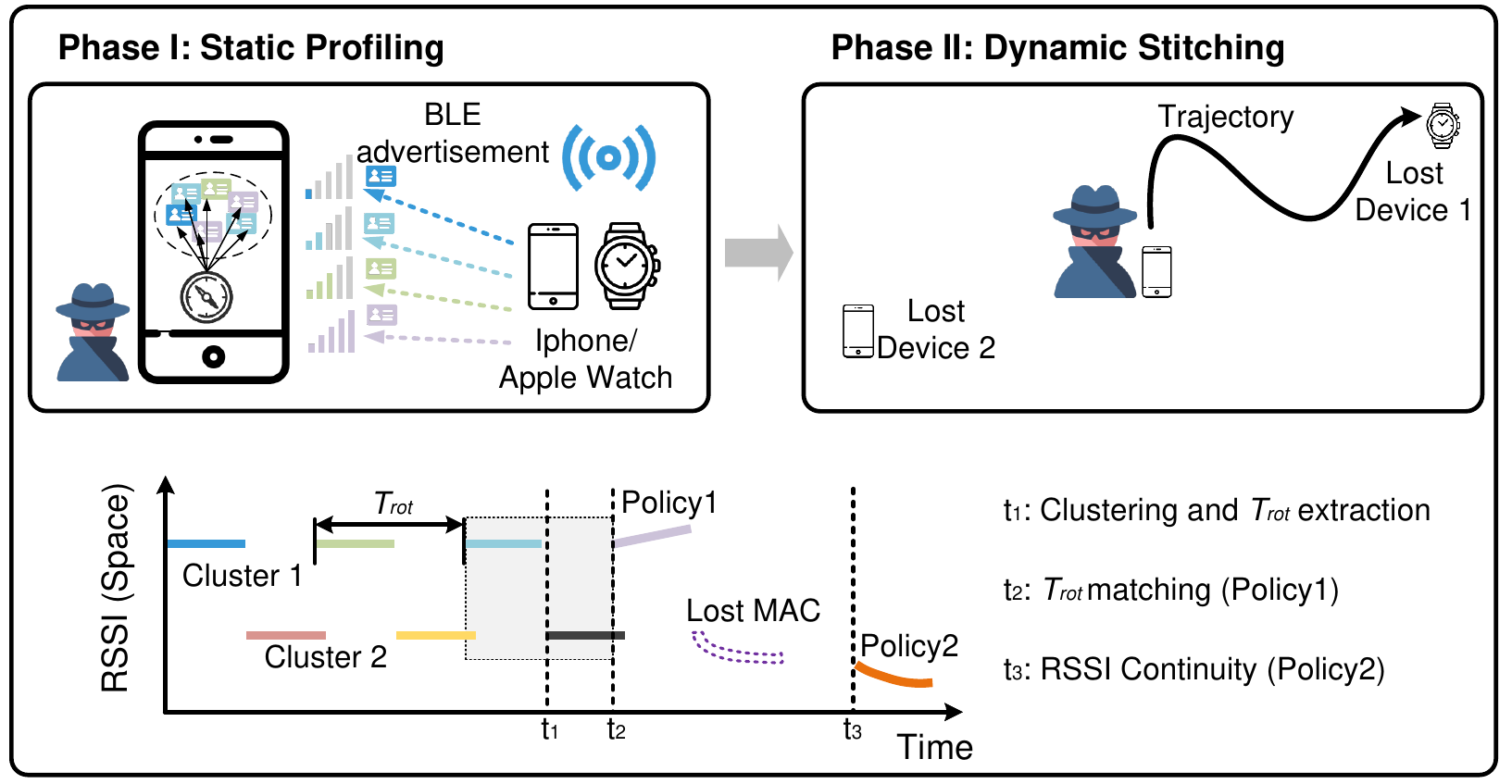}
    \caption{The two-phase identity stitching algorithm. In Phase I (Static Profiling, $t < t_1$), spatial stratification of BLE signals is used to cluster MAC addresses and extract the rotation period ($T_{rot}$). In Phase II (Dynamic Stitching, $t > t_1$), a dual-policy approach stitches new identities. At $t_2$, a new MAC address is uniquely associated using Policy 1 ($T_{rot}$ matching). At $t_3$, Policy 2 (RSSI Continuity) is used to break a tie when Policy 1 can not produce a result.}
    \label{fig:stitching_process}
\end{figure}

\subsection{Level 3: De-anonymization via Spatial-Temporal Clustering}
\label{sec:stage3_design}

In advanced defense scenarios, devices may employ high-frequency MAC address rotation to evade tracking. When multiple lost devices co-exist, the broadcast channel degenerates into a chaotic stream of short-lived MAC addresses. To maintain a lock on one target, as illustrated in Figure~\ref{fig:stitching_process}, we adopt a simple \textit{spatial-temporal clustering} procedure: in Phase~I (\textit{Static Profiling}), the attacker stands still and groups packets into per-device clusters using their average RSSI as a spatial cue, and estimates each cluster’s rotation period $T_{rot}$ as a temporal cue; in Phase~II (\textit{Dynamic Stitching}), the attacker moves and re-associates new MAC addresses to the target by checking whether their rotation time matches the learned $T_{rot}$ and whether their RSSI evolves continuously along the trajectory.

\subsubsection{Phase I: Static Profiling (Learning the Temporal Rhythm)}

In this phase, the attacker stands still and collects BLE advertisements for a fixed time window (e.g, 60\,s in this paper).
The goal is to first cluster the MAC addresses based on their RSSI spatial signatures, and then extract each cluster's rotation period $T_{rot}$.

\textbf{1. Spatial Separation via RSSI Clustering.}
When the attacker does not move, the distances to different lost devices are approximately constant, so their average RSSI levels are also approximately stable but differ across devices (e.g., one device at $-60$ dBm, another at $-80$ dBm).  As shown in the lower panel of Figure~\ref{fig:stitching_process} for $t < t_1$, packets from different devices form roughly separated bands along the RSSI axis rather than a single mixed cloud.
We fit a Gaussian Mixture Model (GMM) to the RSSI values,
and assign each packet to one cluster. In this way, packets are grouped into per-device streams according to their typical RSSI level.

\textbf{2. Estimating the MAC address rotation period.}
Within each RSSI cluster, we then look at the timestamps of MAC address changes. For a given device, an old MAC address disappears at time $t_{\text{death}}$ and a new MAC address appears at $t_{\text{birth}}$, forming repeated \textit{death--birth} pairs over time. By measuring the periods between consecutive MAC address lifetimes in the same cluster, we estimate that device's rotation period $T_{rot}$. This per-device $T_{rot}$ is the temporal feature that Phase II will use to re-associate new MAC addresses during navigation.

\begin{figure}[t]
    \centering
    \begin{subfigure}[b]{.48\linewidth}
        \centering
        \includegraphics[width=\linewidth]{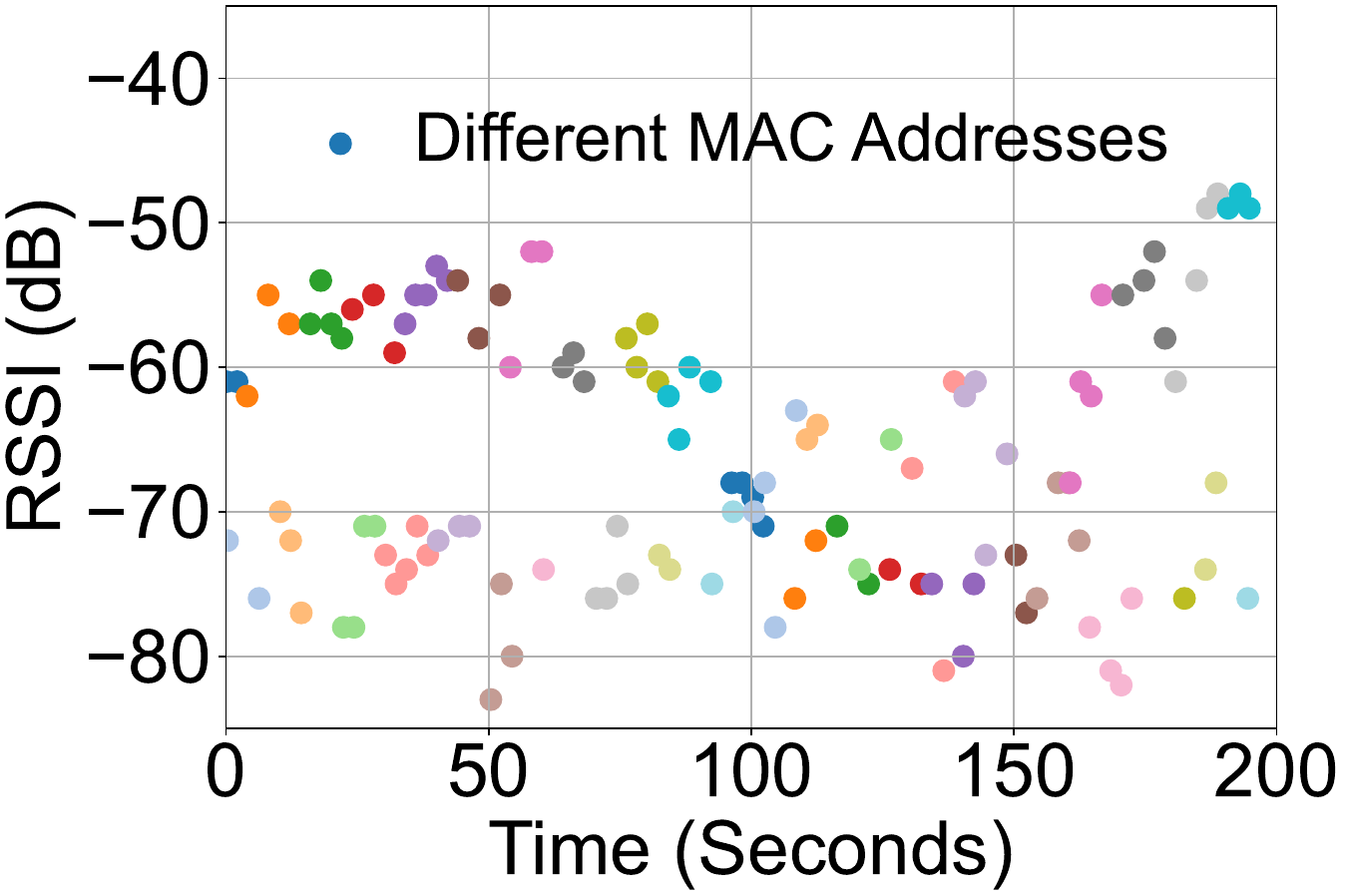}
        \caption{Raw RSSI measurements with MAC address rotation.}
        \label{fig:mac_shuffle_verificationa}
    \end{subfigure}
    \hfill 
    \begin{subfigure}[b]{.48\linewidth}
        \centering
        \includegraphics[width=\linewidth]{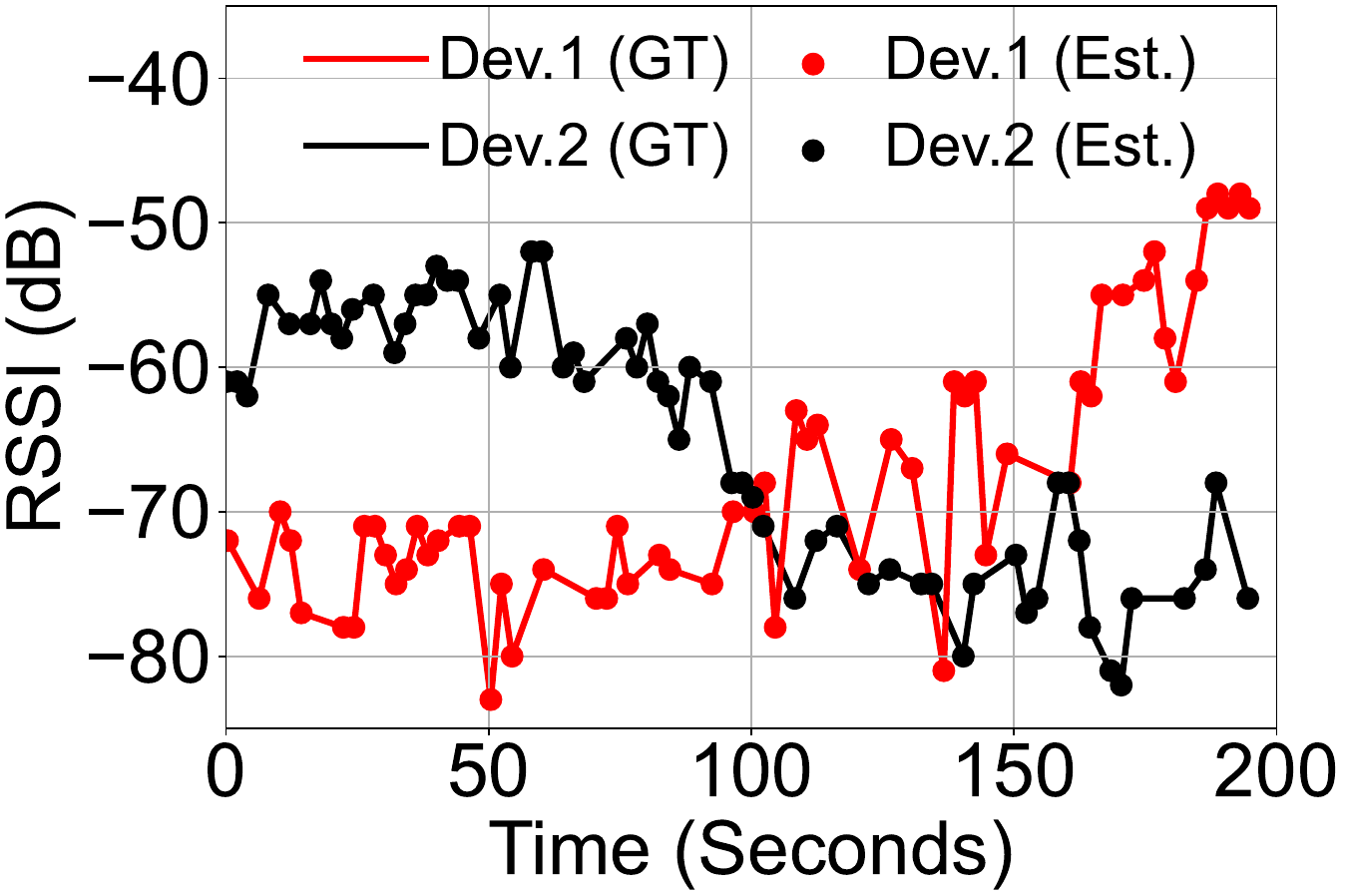}
        \caption{The disentangled RSSI trajectories.}
        \label{fig:mac_shuffle_verificationb}
    \end{subfigure}
    \caption{\textbf{Performance of Identity Stitching.} (a) Raw data appears as fragmented segments due to MAC address randomization. (b) The algorithm disentangles the chaotic stream, recovering continuous trajectories for the two devices.}
    \label{fig:mac_shuffle_verification}
\end{figure}

\subsubsection{Phase II: Dynamic Stitching (Tracking During Navigation)}
After $T_{rot}$ is learned, the attacker starts moving and performs navigation. As the target's MAC address keeps rotating, the key question becomes: when the current MAC address disappears, which newly appearing MAC address belongs to the same physical device.
We treat this as a re-association problem that first uses \textit{temporal alignment} (Policy~1) and, only when necessary, \textit{RSSI continuity} (Policy~2), as shown in Figure~\ref{fig:stitching_process} for $t > t_1$.

\textbf{Policy 1: Temporal alignment.}
Given the learned $T_{rot}$, the system predicts roughly when the next MAC address rotation should occur. When the current MAC address disappears at time $t_{\text{death}}^{\text{curr}}$, we search for new MAC addresses whose first appearance time $t_{\text{birth}}^{\text{new}}$ falls into a small window around this time:
\begin{equation}
    \left| t_{\text{birth}}^{\text{new}} - t_{\text{death}}^{\text{curr}} \right| \le \delta_{\text{sync}}.
\end{equation}
Only MAC addresses satisfying this condition are treated as candidates. In most cases, exactly one candidate satisfies the timing constraint, and we directly assign this new MAC address to the same cluster as the previous one (illustrated at $t_2$ in Figure~\ref{fig:stitching_process}).

\textbf{Policy 2: RSSI continuity.}
In dense environments, multiple new MAC addresses may pass the temporal alignment test (Policy 1) due to packet loss or slight timing variations. we further check which candidate has an RSSI value most consistent with the expected RSSI of the tracked device at that moment. Using the Level~2 tracker, we predict the expected RSSI $\hat{r}_t$ at the time of the handover and select
\begin{equation}
    m_{\text{next}} = \arg\min_{m \in \text{Candidates}} \left| \text{RSSI}_m - \hat{r}_t \right|.
\end{equation}
Intuitively, while the MAC address changes abruptly, the physical signal strength should change smoothly along the user's path, so the correct successor MAC address is the one whose RSSI best follows the previous RSSI trend (illustrated at $t_3$ in Figure~\ref{fig:stitching_process}).

By first using rotation timing and then applying RSSI continuity only when necessary, the algorithm re-links fragmented MAC address segments into a single, continuous trajectory for each physical device.

\noindent
\textbf{Preliminary Verification.}
To validate the efficacy of our stitching algorithm, we conducted a preliminary experiment involving two separate devices performing MAC address rotating at different period.
Figure~\ref{fig:mac_shuffle_verification} visualizes the results. In the raw observation (Figure~\ref{fig:mac_shuffle_verificationa}), the signal landscape is fragmented into numerous disjoint segments (indicated by varying colors), making it impossible to track any specific target.
However, after applying our spatial-temporal clustering (Figure~\ref{fig:mac_shuffle_verificationb}), the chaotic stream is successfully disentangled into two coherent trajectories (Red and Black lines).
Despite the frequent identity rotations, the algorithm correctly stitches the temporal handovers, recovering the continuous RSSI trends for both devices. This result confirms that our method can effectively neutralize MAC address randomization defenses in real-world dynamic scenarios.

Moreover, we validated the end-to-end system in a multi-device scenario with MAC address rotation, conducted within the same environment (Figure~\ref{fig:user_study}d). As depicted in Figure~\ref{fig:all_three}c, two devices were active, frequently changing their MAC addresses (represented by different shapes). Despite this, our \textit{spatial-temporal clustering} algorithm successfully disentangled the chaotic signals, correctly identified the target (Lost device 1), and guided the user to its location. This demonstrates the practical effectiveness of our spatial-temporal clustering in neutralizing MAC address randomization defenses.

\section{Performance Evaluation}
\label{sec:evaluation}
In this section, we evaluate the performance of \textit{Snatcher} through extensive experiments. We first introduce the implementation details and experimental setup. Then, we evaluate the navigation performance of each level, followed by the end-to-end navigation performance across different commercial Apple devices and diverse scenarios.

\subsection{Implementation \& Experimental Setup}
\label{subsec:implementation}

\begin{figure}[!ht]
    \centering
    \includegraphics[width=\linewidth]{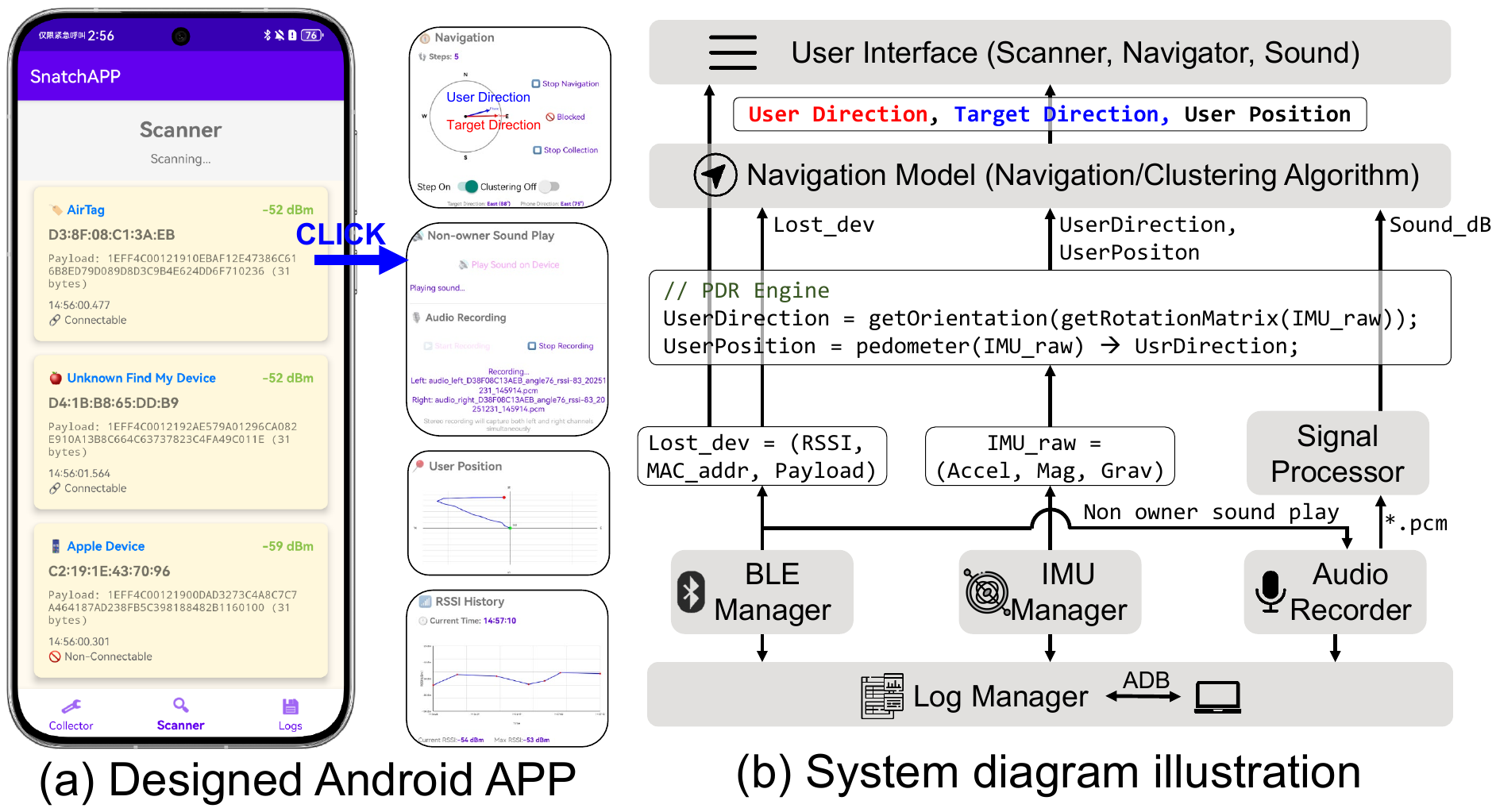}
    \caption{Illustration of the implemented \textit{Snatcher}.}
    \label{fig:implementation}
\end{figure}

\subsubsection{Implementation of \textit{Snatcher}}

\textit{Snatcher} is implemented as an integrated Android application \cite{anonymous_snatcher_software_2026} that realizes the three-level attack framework. Figure~\ref{fig:implementation} illustrates the application's user interface and its underlying system architecture.

The application's UI, shown in Figure~\ref{fig:implementation}a, is designed for efficient device discovery and navigation. The main screen provides a \textit{Scanner} that discovers nearby lost Apple devices by filtering BLE advertisements with the Find My features that was detailed in Section~\ref{sec:primer}. To maximize detection performance, the scanning mode is configured to \texttt{SCAN\_MODE\_LOW\_LATENCY} using Android's standard Bluetooth API~\cite{android_scan_settings}. Tapping a device entry transitions the user to a multi-panel navigation interface that provides real-time feedback through several modules: a \textit{Navigation} panel displaying the user's orientation versus the target's direction, a \textit{Non-owner Sound Play} panel to trigger acoustic signals, a \textit{User Position} panel visualizing the reconstructed trajectory, and an \textit{RSSI History} panel showing signal strength trends.

As illustrated in Figure~\ref{fig:implementation}b, the system's backend is modularized to correspond with the three attack levels, leveraging standard Android APIs and open-source libraries:
\begin{itemize}
    \item \textbf{Level 1 (Acoustic):} For accessories like AirTags, the \texttt{Lost Device Connector} establishes a BLE connection to trigger sound using anti-stalking APIs from \texttt{AirGuard}~\cite{10.1145/3507657.3528546}. The \texttt{Audio Recorder}, via the \texttt{AudioRecord} API~\cite{android_audio_record}, captures incoming audio. This data is then processed by the \texttt{Signal Processor}, which applies a 2.6–2.8\,kHz band-pass filter and uses cross-correlation to calculate sound signal strength and estimate the sound source direction.

    \item \textbf{Level 2 (RSSI-IMU):} To locate silent devices, the \texttt{BLE Manager} collects RSSI data while the \texttt{IMU Manager} gathers sensor data. A Pedestrian Dead Reckoning (PDR) \cite{rashid2015dead} engine reconstructs the user's movement using the \texttt{getOrientation} and \texttt{getRotationMatrix} API~\cite{android_sensor_manager} for heading and Google's open source \texttt{pedometer} module~\cite{google_simple_pedometer} for step counting. Both data streams serve as input to the \texttt{Navigation Model}, which performs a fusion process to compute a robust heading vector toward the target.

    \item \textbf{Level 3 (Clustering):} To counter MAC address randomization, the \texttt{Navigation Model} incorporates the spatial-temporal clustering algorithm to stitch together fragmented identities. For evaluation, a \textit{mimicked AirTag} was implemented by modifying the \texttt{OpenHaystack}~\cite{openhaystack} firmware on an ESP32 board, allowing programmable MAC address rotation period (10–60\,s).

    \item \textbf{Logging and Data Access}. A centralized \texttt{Log Manager} continuously records all sensor data streams (BLE, IMU, audio). These logs are accessible via the Android Debug Bridge (ADB), facilitating offline analysis and ensuring the reproducibility of our experiments.
\end{itemize}

\begin{figure}[t]
    \centering
    \includegraphics[width=0.5\linewidth]{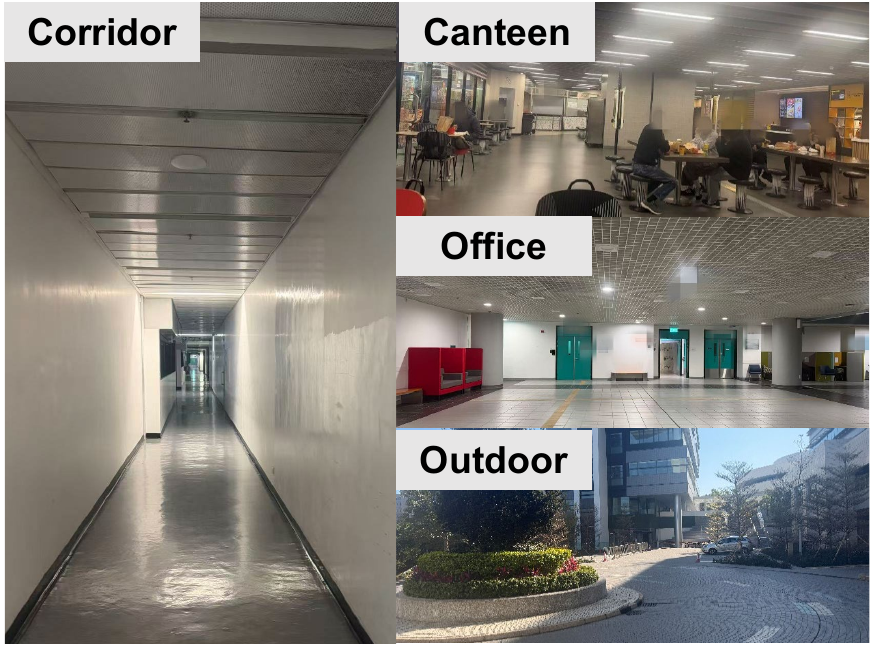}
    \caption{Illustration of indoor and outdoor evaluated scenarios.}
    \label{fig:scenarios}
\end{figure}

\subsubsection{\blue{Experimental Setup}}
\blue{We conduct extensive experiments in various real-world environments, including an indoor office, a corridor, a canteen, and an outdoor parking lot, as shown in Figure~\ref{fig:scenarios}. Our evaluation is structured into two main parts: a level-specific analysis to assess the core performance of each attack stage (Subsection~\ref{subsec:level}), and an end-to-end evaluation to measure \textit{Snatcher}'s overall effectiveness (Subsection~\ref{subsec:end2end}).
For rigorous evaluation, we repeated each attack level and the end-to-end run over 20 times, where experimenters strictly follow on-screen guidance.}

For the level-specific evaluation, we focus on validating the fundamental capabilities and limitations of each navigation algorithm. Our primary metric is the navigation success rate, which reflects the algorithm's autonomous performance. A navigation task is deemed failed if the algorithm, without human intervention, continuously guides the user away from the target for more than 300\,s.

\begin{itemize}
    \item For Level 1, we evaluate the impact of ambient noise on the success rate by testing in quiet and noisy environments. We also analyze the relationship between signal strength, connection overhead, and distance to determine its practical operational range.

    \item For Level 2, we compare the success rate of our fused-gradient algorithm against a baseline short-term gradient method to demonstrate its robustness. Similarly, we analyze signal strength and scanning overhead versus distance to define its effective range.

    \item For Level 3, we assess the success rate of spatial-temporal assisted navigation when faced with two or three co-located devices rotating their MAC addresses. Additionally, we measure the clustering accuracy of our \textit{spatial-temporal clustering} as a function of the MAC address rotation period.
\end{itemize}

For the end-to-end navigation evaluation, we assess \textit{Snatcher}'s practical performance by conducting navigation tasks with multiple commercial Apple devices, such as AirTags, iPhones, and Apple Watches, across the various scenarios mentioned above. The key metrics for this evaluation are the total navigation time and navigation distance required to successfully find the target. This holistic assessment demonstrates \textit{Snatcher}'s real-world applicability and effectiveness.

\subsection{Level-specific evaluation}
\label{subsec:level}

In this subsection, we evaluate the lost device navigation performance for acoustic and RSSI based navigation, focusing on the navigation success rate and maximum navigation distance analysis.

\subsubsection{Level 1}

\begin{figure}[t]
    \centering
    \begin{subfigure}[t]{0.48\linewidth}
        \centering
        \includegraphics[width=\linewidth]{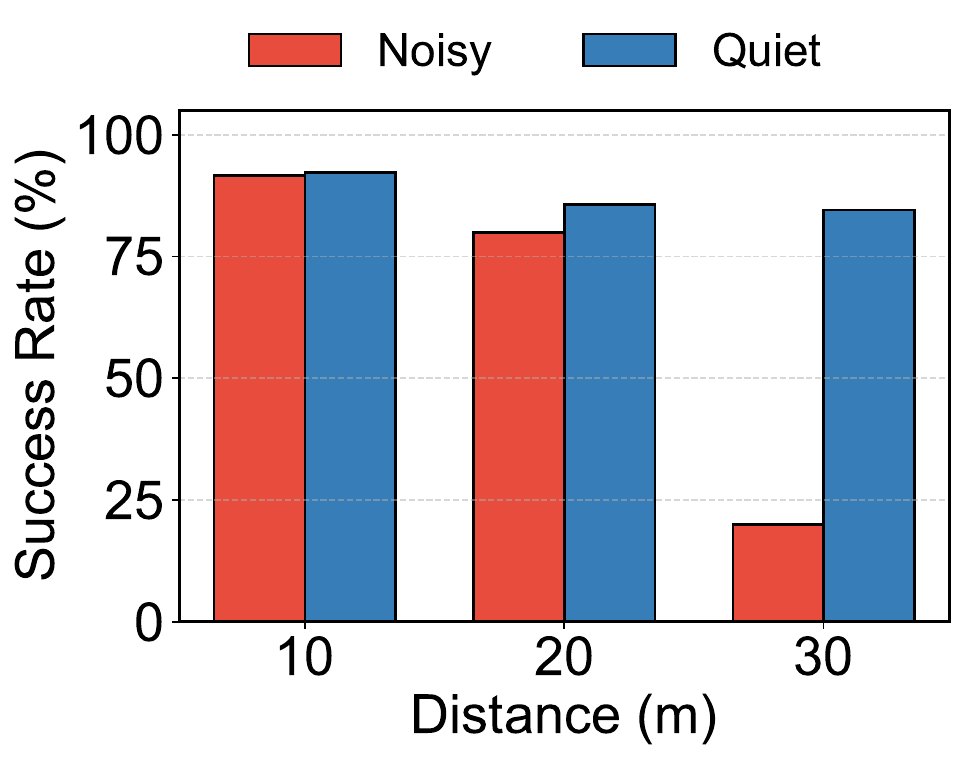}
        \caption{Acoustic navigation success rate vs. distance under noisy and quiet environment.}
        \label{fig:nav_stage1_success_rate}
    \end{subfigure}
    \hfill
    \begin{subfigure}[t]{0.48\linewidth}
        \centering
        \includegraphics[width=\linewidth]{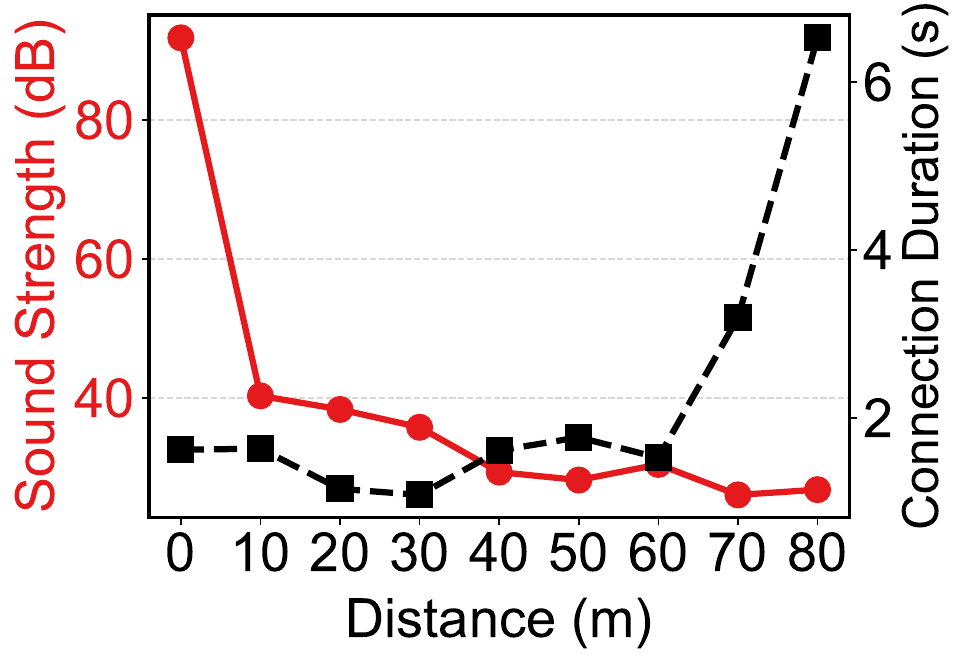}
        \caption{Average sound strength (dB) and connection duration (s) versus distance (m).}
        \label{fig:sound_connection_vs_distance}
    \end{subfigure}
    \caption{Performance of acoustic navigation and connection versus distance. 
    (a) Navigation success rate. (b) Sound strength and connection duration.}
    \label{fig:combined_nav}
    \vspace{-2mm}
\end{figure}

Figure~\ref{fig:combined_nav} provides a quantitative comparison of the acoustic navigation success rate as a function of distance in both quiet and noisy environments. In the quiet office setting, the system demonstrates robust and stable performance. At a distance of 10\,m, it achieves a high success rate of approximately 92\%, which only slightly decreases to around 86\% at 20\,m and remains high at about 85\% even at 30\,m. 
In contrast, the system's performance in the noisy canteen environment degrades significantly with distance. While it starts with a respectable success rate of about ~91\% at 10\,m, this rate drops sharply to around ~80\% at 20\,m and plummets to a mere ~20\% at 30\,m, due to signal obscuration by background noise.

Furthermore, we analyze the relationship between distance, signal quality, and connection overhead, as presented in Figure~\ref{fig:sound_connection_vs_distance}. Our measurements show that the non-owner sound play can be triggered via the GATT interaction flow at distances up to 80\,m.
However, two key factors limit the effective navigation range. First, the acoustic signal strength (red line) decays rapidly, falling below 40\,dB beyond 30\,m, making it susceptible to ambient noise interference. Second, the connection duration (black dashed line) increases significantly beyond 60\,m, with exponential growth that makes the iterative navigation strategy inefficient (up to 7.5\,s).

\subsubsection{Level 2}

\begin{figure}[t]
    \centering
    \begin{subfigure}[t]{0.48\linewidth}
        \centering
        \includegraphics[width=\linewidth]{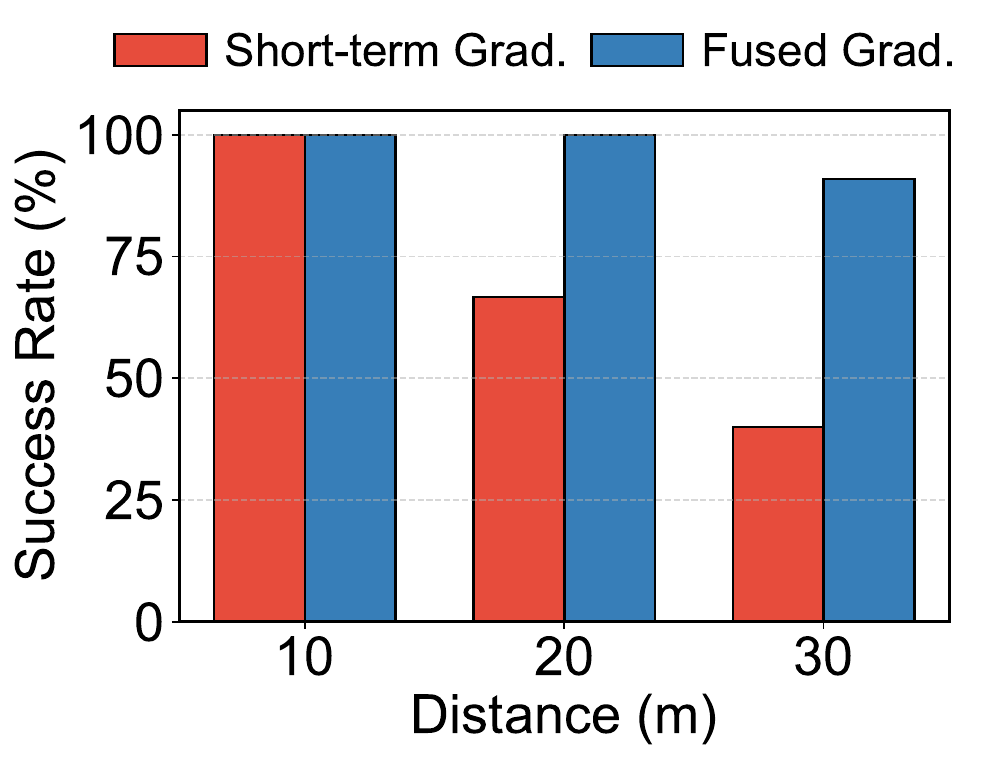}
        \caption{RSSI-IMU navigation success rate vs. distance using short-term gradient and fused gradient.}
        \label{fig:nav_stage2_success_rate}
    \end{subfigure}
    \hfill
    \begin{subfigure}[t]{0.48\linewidth}
        \centering
        \includegraphics[width=\linewidth]{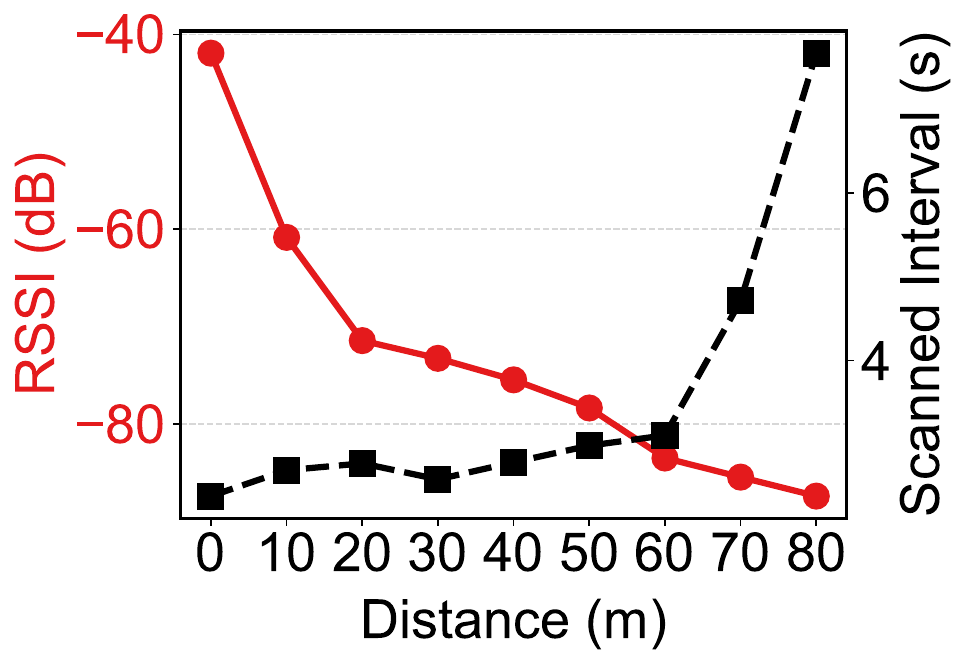}
        \caption{Average RSSI (dB) and scanned advertisement interval (s) versus distance (m).}
        \label{fig:rssi_adv_interval_vs_distance}
    \end{subfigure}
    \caption{Performance of RSSI-IMU navigation and signal behavior versus distance.
    (a) Navigation success rate. (b) RSSI and scanned advertisement interval.}
    \label{fig:combined_rssi}
    \vspace{-2mm}
\end{figure}

\begin{figure}[t]
    \centering
    \begin{subfigure}[t]{0.48\linewidth}
        \centering
        \includegraphics[width=\linewidth]{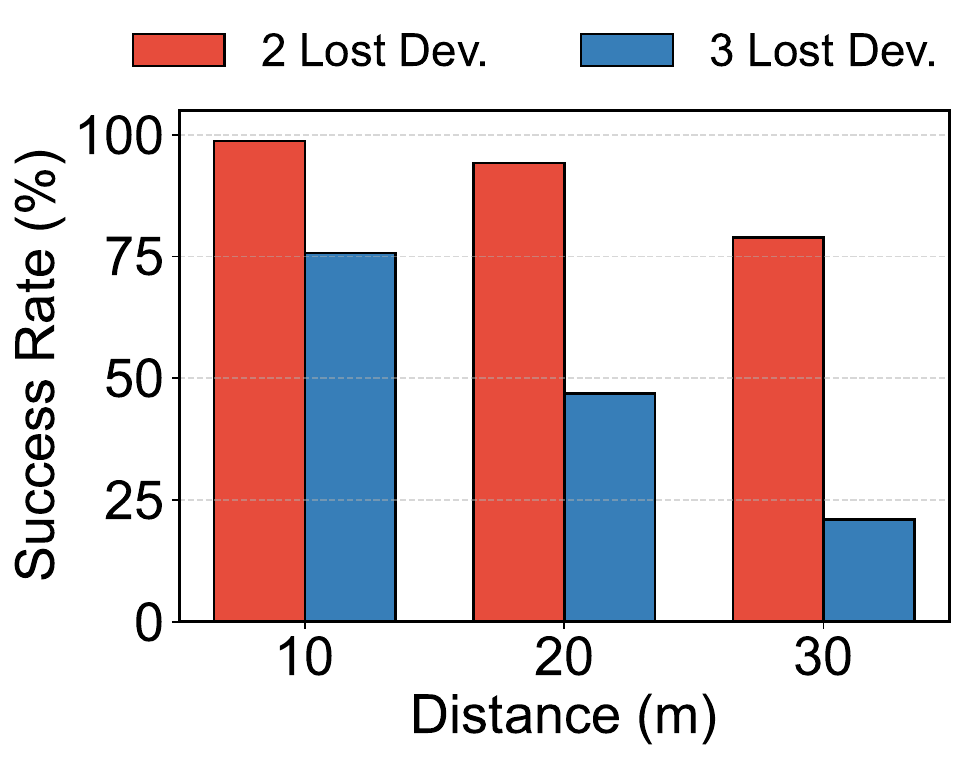}
        \caption{Spatial-temporal assisted RSSI-IMU navigation success rate vs. distance under 2 and 3 lost-device MAC address rotation (30\,s period) scenarios.}
        \label{fig:nav_stage3_success_rate}
    \end{subfigure}
    \hfill
    \begin{subfigure}[t]{0.48\linewidth}
        \centering
        \includegraphics[width=\linewidth]{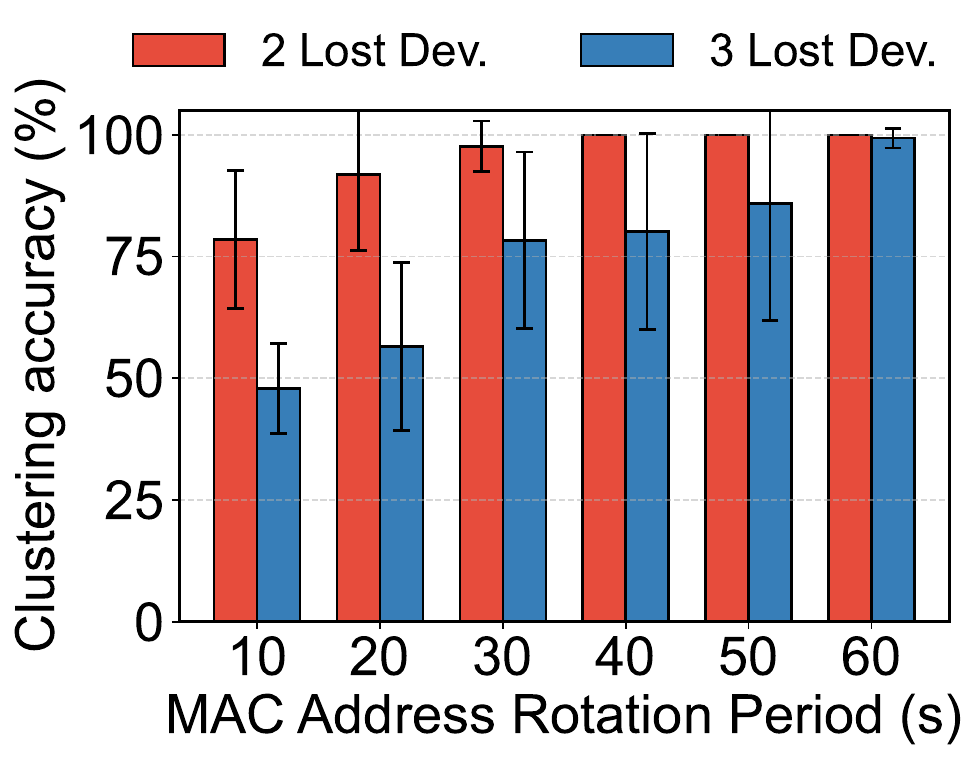}
        \caption{Accuracy of spatial-temporal clustering for 2 or 3 lost devices periodically rotating MAC addresses every 10–60\,s.}
        \label{fig:stage3_accuracy}
    \end{subfigure}
    \caption{Performance of our spatial-temporal assisted algorithm under MAC address rotation. We evaluate with two and three lost devices, representing realistic scenarios where few devices are likely to be co-located due to BLE's short communication range. (a) Navigation success rate. (b) Clustering accuracy versus rotation period.}
    \label{fig:combined_stage3}
\end{figure}

\begin{figure*}[t]
    \centering
    \begin{subfigure}[t]{0.45\linewidth}
        \includegraphics[width=\linewidth]{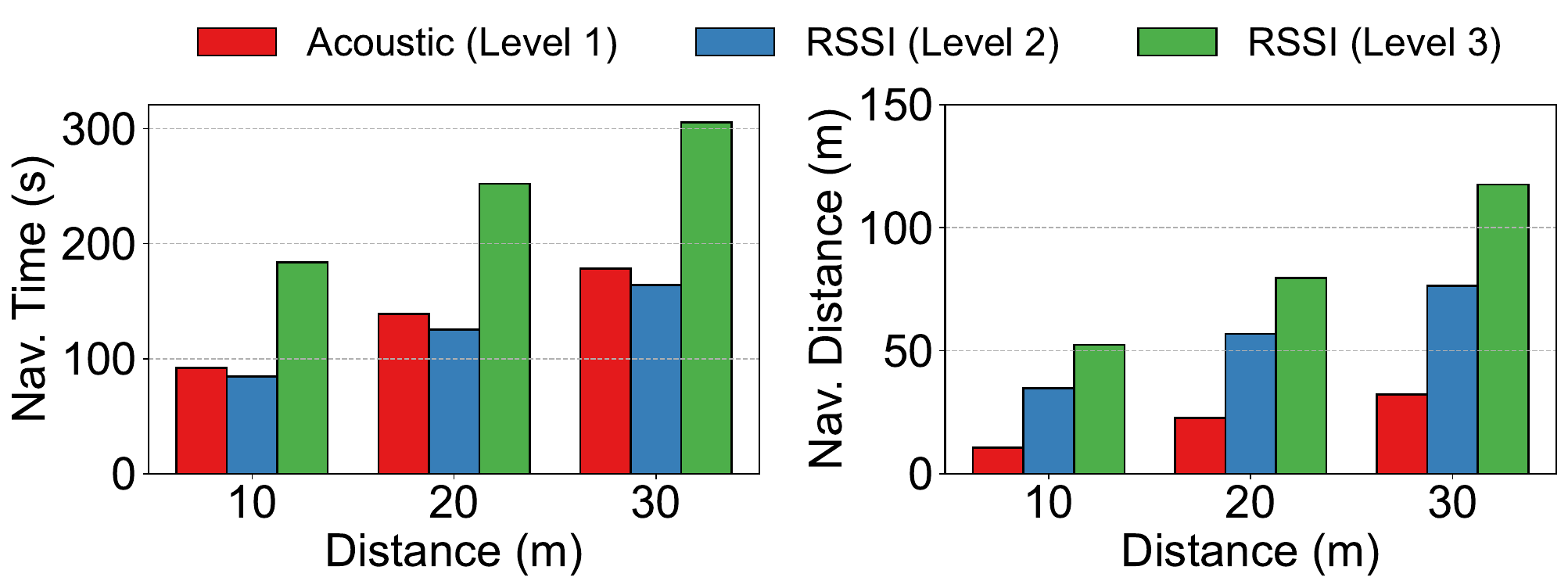}
        \caption{Indoor office}
    \end{subfigure}
    \qquad
    \begin{subfigure}[t]{0.45\linewidth}
        \includegraphics[width=\linewidth]{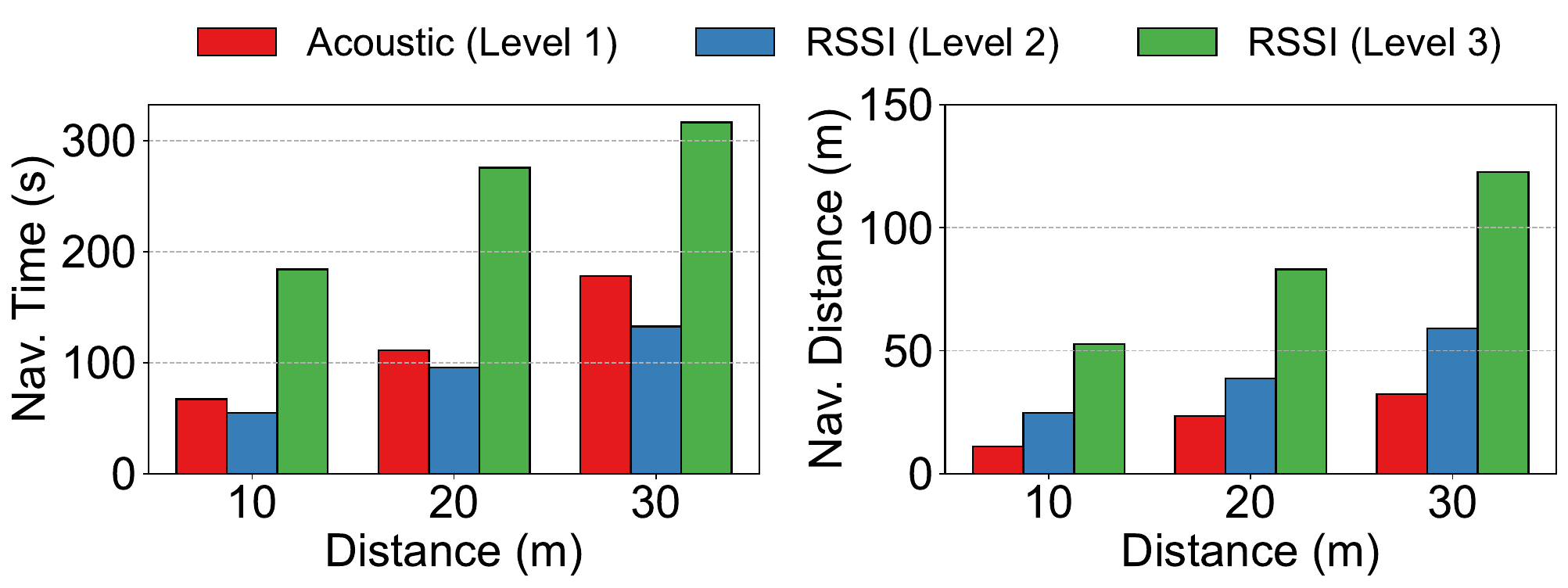}
        \caption{Indoor corridor}
    \end{subfigure}    
    \\
    \begin{subfigure}[t]{0.45\linewidth}
        \includegraphics[width=\linewidth]{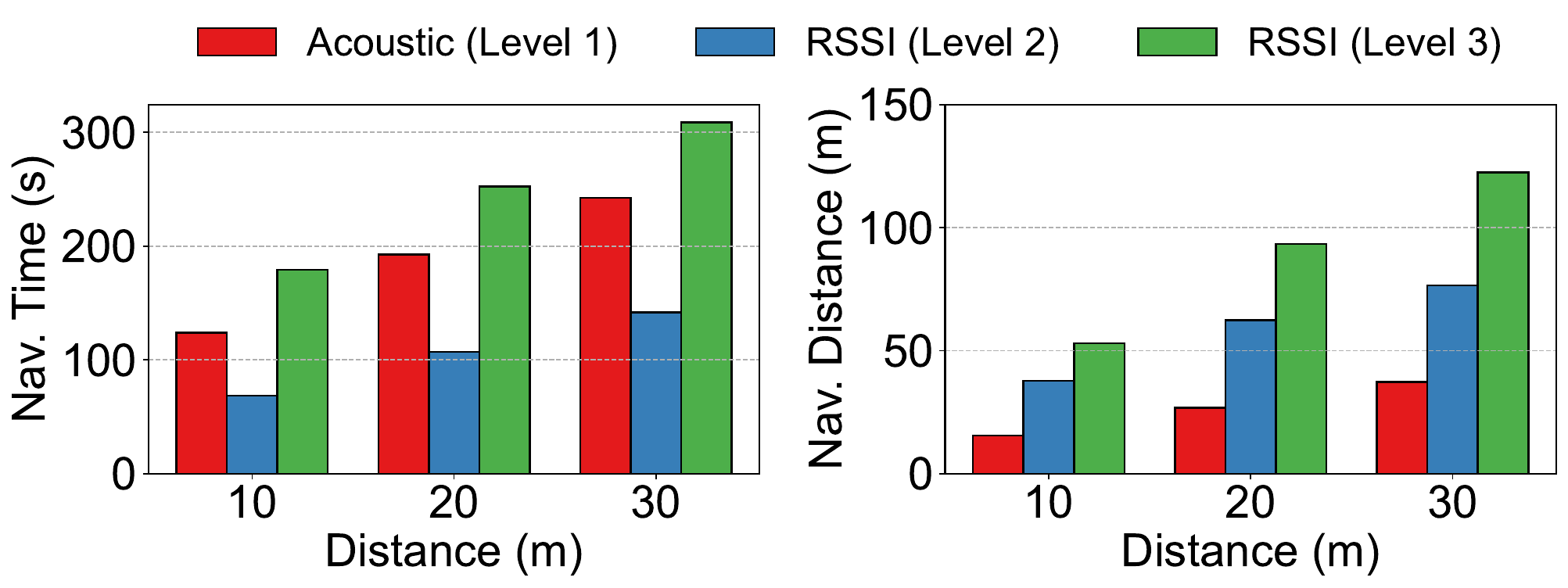}
        \caption{Indoor canteen}
    \end{subfigure}
    \qquad    
    \begin{subfigure}[t]{0.45\linewidth}
        \includegraphics[width=\linewidth]{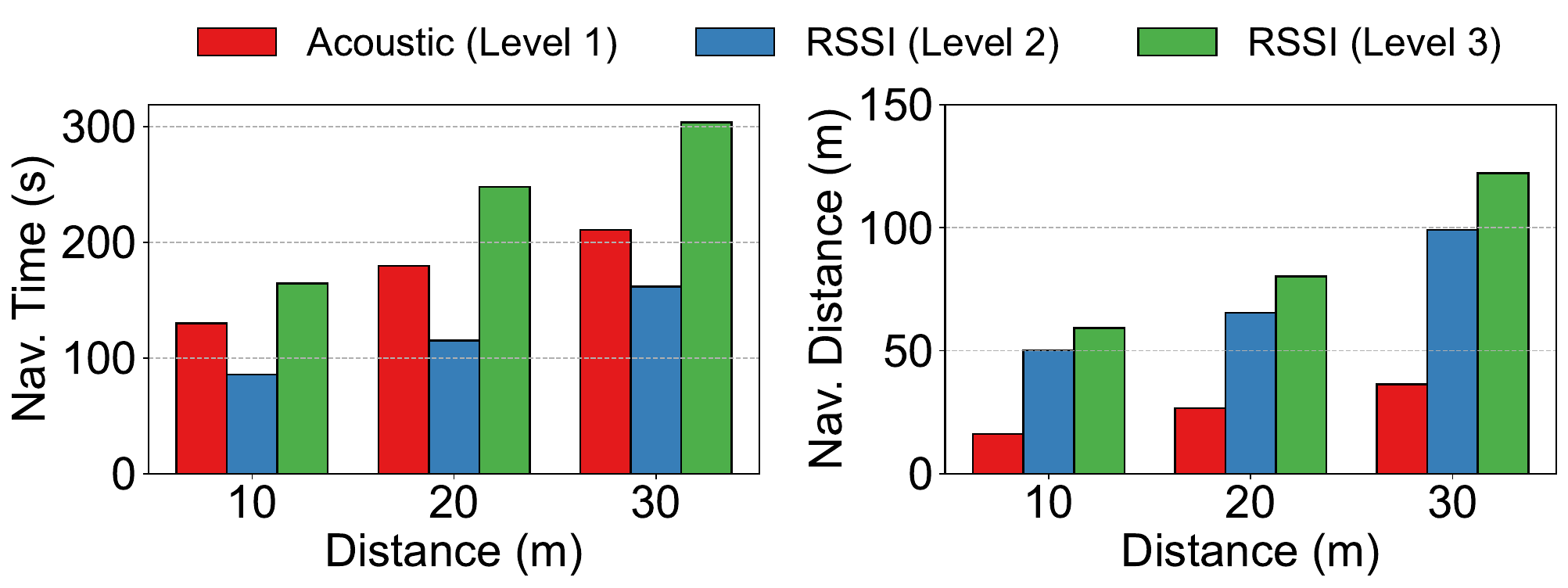}
        \caption{Outdoor parking lot}
    \end{subfigure}   
    \caption{Evaluation of average navigation time and distance for different scenarios. For Level~3, the mimicked MAC address rotation period is 30\,s with two lost devices.}
    \label{fig:evaluation_disance_time}
    \vspace{-3mm}
\end{figure*}

We now evaluate the RSSI-IMU navigation algorithm for silent devices, focusing on its success rate and effective operational range. The performance is detailed in Figure~\ref{fig:combined_rssi}. 
Specifically, Figure~\ref{fig:nav_stage2_success_rate} compares the success rate of two RSSI-IMU navigation algorithms: a short-term RSSI gradient method and our proposed fused-gradient approach. The proposed method, which fuses short-term and global trajectory trends, demonstrates superior performance. 

As shown in Figure~\ref{fig:nav_stage2_success_rate}, the baseline short-term method's success rate drops sharply from 100\% at 10\,m to only 40\% at 30\,m. While our fused-gradient algorithm maintains a 100\% success rate up to 20\,m and sustains a high 90\% success rate at 30\,m. The performance disparity stems from how each algorithm handles signal instability. The short-term method is highly susceptible to natural signal fluctuations and multipath effects. This leads to frequent misjudgments and a \textit{ping-pong} effect, where the user is guided back and forth, especially at greater distances. Our fused-gradient algorithm mitigates these issues by leveraging the global trajectory. This allows it to stabilize direction updates against local variations and bias exploration towards the global gradient descent, thus avoiding local optima.

Figure~\ref{fig:rssi_adv_interval_vs_distance} illustrates the signal reception characteristics and associated overhead. A key metric for navigation overhead is the scanned advertisement interval.
Consistent with the Find My network's ~2-second broadcast period \cite{AppleFindMySpec2020} and standard BLE random advertisement delays \cite{BluetoothSIGSpecV6}, the measured interval remains stable at approximately 2\,s for distances up to 60\,m. 
We attribute this surge to significant packet loss caused by channel fading, which likely reaches 50\% at 70\,m and 70\% at 80\,m.
This high overhead creates a critical bottleneck: it provides insufficient data samples for the algorithm to compute a reliable gradient. This data scarcity not only compromises the accuracy of the navigation but also forces longer wait times to accumulate enough samples, further prolonging the overall process.

\subsubsection{\blue{Level 3}}

\blue{
We evaluated our spatial-temporal clustering algorithm's performance against an emulated advanced defense scenario. This scenario featured high-frequency MAC address rotation, implemented using the aforementioned ESP32-based \textit{mimicked AirTags}.
Figure~\ref{fig:nav_stage3_success_rate} presents the end-to-end navigation success rate when assisted by the clustering algorithm, with a fixed MAC address rotation period of 30\,s. 
}
\blue{In the two-device scenairio, the system maintains a comparable success rate, achieving ~80\% at 30\,m. In the more challenging three-device scenario, the success rate drops to around ~20\% at 30\,m. The sharp drop stems from error accumulation during joint identification and navigation: a single misidentified MAC address diverts the attacker to the wrong target, causing "ping-pong" navigation paths that ultimately prolong or fail the attacking attempt.}

Figure~\ref{fig:stage3_accuracy} shows the clustering accuracy of our identity stitching algorithm. 
The clustering accuracy is calculated as the proportion of lost device MAC addresses that are correctly clustered out of the total number of MAC addresses received from those devices.
In the two-device scenario, the algorithm achieves near-perfect accuracy (>98\%) for rotation periods of 30\,s or longer. Even with a rapid 10-second rotation period, the accuracy remains high at approximately 80\%. When the complexity increases to three devices, the accuracy shows a slight decline but remains robust, maintaining approximately over 80\% accuracy for intervals of 30\,s and above.
For the remainder of our end-to-end evaluations, we will use the setting of two lost devices with a 30-second rotation period as the representative scenario for Level 3 attacks.

\subsection{End-to-end navigation evaluation}
\label{subsec:end2end}

In this section, we first evaluate the end-to-end navigation performance across different commercial Apple devices in various scenarios, based on the previously determined experiment setup. Then, we analyze the navigation time and distance overhead for each level across different scenarios.

\subsubsection{\blue{Navigation performance across different commercial Apple devices}}

\begin{table}[t]
\centering
\caption{\blue{Navigation performance of Apple Accessory (AirTag \& AirPods, top) and Apple Device (iPhone \& Apple Watch, bottom). Mean $\pm$ std over all trials; path efficiency = target straight-line distance / actual walked distance (closer to 1 is better).}}
\label{tab:summary_nav_commerical_device}
\small
\setlength{\tabcolsep}{4pt}
\begin{tabular}{@{}ccccc@{}}
\hline
& Dist. & Nav. Time & Nav. Dist. & Path Eff. \\ 
\hline
\multirow{3}{*}{\begin{tabular}[c]{@{}c@{}}Apple Accessory\\(AirTag \& AirPods)\end{tabular}}
  & 10 m & 103.2 $\pm$ 34.0 s & 13.3 $\pm$ 5.0 m & 0.81 $\pm$ 0.19 \\
  & 20 m & 155.7 $\pm$ 48.7 s & 24.9 $\pm$ 4.1 m & 0.82 $\pm$ 0.12 \\
  & 30 m & 202.4 $\pm$ 54.2 s & 34.5 $\pm$ 5.7 m & 0.89 $\pm$ 0.14 \\ 
\hline
\multirow{3}{*}{\begin{tabular}[c]{@{}c@{}}Apple Device\\(iPhone \& Watch)\end{tabular}}
  & 10 m & 73.4 $\pm$ 27.1 s & 36.8 $\pm$ 17.3 m & 0.34 $\pm$ 0.17 \\
  & 20 m & 110.9 $\pm$ 40.4 s & 55.9 $\pm$ 24.4 m & 0.43 $\pm$ 0.19 \\
  & 30 m & 150.0 $\pm$ 42.8 s & 77.8 $\pm$ 28.6 m & 0.44 $\pm$ 0.17 \\ 
\hline
\end{tabular}
\end{table}

\blue{
We evaluate \textit{Snatcher}'s performance on two primary categories of commercial Apple products: Apple Accessories (AirTags and AirPods), which are targeted by our Level 1 acoustic navigation, and Apple Devices (iPhones and Apple Watch), targeted by our Level 2 silent navigation. \blue{Table~\ref{tab:summary_nav_commerical_device} summarizes the navigation time, navigation distance, and path efficiency (calculated by straight-line distance/actual walked distance) required to locate these devices from starting points of 10\,m, 20\,m, and 30\,m}. 
}

\blue{
While the Level 1 acoustic method results in a more direct navigation path, it requires more time than the Level 2 RSSI-based approach. This is due to the overhead of the acoustic feedback loop. 
}
\blue{The \textit{Initial Direction Finding} and \textit{Step-wise Navigation} strategies require the attacker to repeatedly establish a BLE connection, trigger the sound, and record audio for several seconds at each decision point, which is inherently time-consuming.}

\blue{
For Apple Accessories (AirTag and AirPods), the acoustic-based navigation successfully located the target in all tests, taking an average of 202.4\,s for the 30\,m scenario while maintaining a relatively direct path (34.5\,m walked, path efficiency of 0.89). For Apple Devices (iPhone and Apple Watch), the RSSI-IMU navigation proved faster in terms of time, requiring only 150.0\,s for the same 30\,m distance. However, it resulted in a considerably longer navigation path (77.8\,m) due to the exploratory movements needed to resolve signal ambiguities, yielding a lower path efficiency of 0.44. 
}

\subsubsection{Navigation performance across different scenarios}
We conduct a comprehensive end-to-end evaluation of \textit{Snatcher}'s three attack levels across four distinct scenarios. Figure~\ref{fig:evaluation_disance_time} presents the average navigation time and distance required to locate a target from starting distances of 10\,m, 20\,m, and 30\,m in each environment.

\textbf{Performance Comparison across Levels.}
As shown in the figure, the navigation time and distance for all levels increase as the initial separation from the target grows. Across all scenarios, the relative performance of the three attack levels remains consistent. Level 1 (Acoustic) generally results in a shorter, more direct navigation path but takes longer than Level 2 due to the overhead of its acoustic feedback loop. Level 3 (RSSI with Clustering) consistently requires the most time and distance. This is because it includes the overhead of Level 2, plus an initial static profiling period of at least 60 seconds to learn the MAC address rotation rhythm. Furthermore, the navigation distance in Level 3 is longer, as the clustering algorithm might occasionally guide the user toward the non-target device before converging on the correct one.

\textbf{Performance Comparison across Scenarios.}
The physical environment significantly impacts navigation efficiency. As shown in Figure ~\ref{fig:scenarios}, our evaluation covers four scenarios chosen to represent distinct real-world challenges.
The substantive differences between these environments are as follows: the indoor corridor offers a simple geometry with limited directional ambiguity; the office and canteen feature complex indoor layouts with physical obstacles; the outdoor parking lot represents an open, unbounded space requiring extensive exploration; and the canteen is further distinguished by high ambient noise levels that interfere with acoustic signals.

The results in Figure~\ref{fig:evaluation_disance_time} show that the indoor corridor scenario yields the shortest navigation times and distances, as the constrained space limits the search to essentially two directions. The indoor office and canteen scenarios show similar overall performance; however, the canteen's high ambient noise and complex layout (pedestrians, tables) increase the time and path length for both Level 1 (due to noise interference) and Level 2/3 (requiring more exploration to navigate obstacles). In the outdoor parking lot, the wide-open space leads to slightly poorer RSSI navigation performance compared to indoor settings. This is because the lack of physical boundaries necessitates deeper and more extensive exploratory movements to establish a reliable signal gradient.

\section{Related Work}
\label{sec:relatedwork}

\noindent\textbf{Security in Crowdsourced Offline-Finding Networks.} Major tech companies have established large-scale Offline-Finding Networks (OFNs), starting with Apple's Find My network~\cite{AppleFindMySpec2020}, and followed by Google's Find My Device Network (FMDN)~\cite{google_fmdn_fhn_spec_v1_3} and Samsung's SmartThings Find~\cite{samsung_smartthings_find_dev_guide}.  
The academic community has since extensively scrutinized these ecosystems.  
Pioneering works first reverse-engineered the proprietary protocols of Apple's, Google's, and Samsung's networks, documenting the foundational mechanisms of device pairing, key management, and location reporting~\cite{openhaystack, heinrich2021can, bottger2025okay, yu2024security}.  
Subsequent research has analyzed their security and privacy from a protocol-level perspective, uncovering a range of vulnerabilities, such as logical flaws in protocol design and the potential for linkage attacks~\cite{liu2025thorough, bottger2025okay, yu2024security}.
Previous work addressed upper-layer privacy vulnerabilities but overlooked physical-layer security risks. Our work fills this gap by demonstrating how adversaries can exploit weakly obfuscated BLE broadcasts to physically locate and steal lost devices.  
  
The network’s unprecedented tracking capability has evolved into a double-edged sword, enabling its abuse such as tracker-based stalking.  
Attackers can surreptitiously place a tag~\cite{stalker, newyorktimes} or a modified commercial Bluetooth device~\cite{openhaystack,chen2025tracking,9833881,burg2022openhaystack} on a target, and then monitor the target’s movements via the Find My network, introducing severe physical safety concerns.  
To mitigate this, recent research~\cite{eldridge2024abuse, GerhardtFGDK25, 10.1145/3507657.3528546, mayberry2023blind,heinrich2024please,shafqat2023track,turk2024stop} and industry standards (e.g., Apple and Google’s IETF document~\cite{IETF}) have heavily pivoted toward anti-stalking mechanisms, such as relying on \textit{non-owner sound play} and long MAC address rotation period policy (e.g., 24 hours) to alert victims~\cite{alert}.  
However, this evolution of defense introduces a new security paradox, which has never been noticed and investigated. We disclose in our work that by weaponizing this defense mechanism as an acoustic side-channel, snatchers can precisely locate concealed assets.

\noindent  
\textbf{Security in BLE.}  
Device tracking is a major research area in Bluetooth security. Although MAC address randomization was introduced to prevent long-term linkage\cite{bluetooth_privacy,bluetooth_randomized_rpa_updates}, studies have shown it can be defeated. At the implementation level, vulnerabilities arise from static identifiers within the BLE specification~\cite{wu2024finding}, asynchronous updates between addresses and payloads~\cite{becker2019tracking}, or static data like sequence counters in advertisements~\cite{celosia2019saving, celosia2020discontinued}. More advanced attacks exploit behavioral side-channels, where security features like the "allowlist" are weaponized to de-anonymize devices~\cite{zhang2022good, ellis2025deanonymizing}.

To bypass these software-level defenses entirely, physical-layer fingerprinting methods exploit unique hardware imperfections. By analyzing radio frequency characteristics such as Carrier Frequency Offset or I/Q imbalance, devices can be tracked even with rotating MAC addresses~\cite{givehchian2022evaluating, li2024ble,10001572,stoian2025augmenting}. However, these approaches typically require specialized Software-Defined Radio (SDR) hardware.
In contrast to these works, we rely solely on Commodity Off-The-Shelf Android smartphones without requiring specialized SDR hardware.  
Moreover, while prior works focus on tracking only, we bridge the gap to physical navigation and theft. We demonstrate that leveraging the spatial-temporal continuity of RSSI, attacker mobility, and timestamps allows for precise localization of the target device.

\vspace{-2mm}
\vspace{-2mm}
\section{Conclusion}
\label{sec:conclusion}

\blue{This paper demonstrates that Apple's Find My network, while designed for user convenience, contains design tradeoffs that inadvertently expose lost devices to physical theft. 
Through our analysis and the implementation of the \textit{Snatcher} framework, we have shown that vulnerabilities in BLE broadcasts, unauthenticated sound triggers, and predictable MAC address randomization can be systematically exploited. Our three-level attack model, utilizing commodity hardware like Android smartphones, successfully locates Apple accessories and devices in real-world scenarios. These findings highlight a significant conflict between the network's goals of privacy, anti-stalking, and physical security, underscoring the urgent need for Apple to implement stronger security measures.}

\balance
\bibliographystyle{ACM-Reference-Format}
\bibliography{refs}

\appendix

\section{Ethics Considerations}

\blue{
All attacks described in this paper were performed on our own devices and mimicked development boards. No user data was accessed or collected during the experiments. 
To mitigate the vulnerabilities demonstrated in this work, we propose three orthogonal countermeasures, which we will report to Apple prior to paper publication:}

\begin{enumerate}
    \item \blue{\textbf{Better control of non-owner sound play.} Instead of permitting unauthenticated external acoustic triggers, separated devices should autonomously decide when to sound alerts. For example, utilizing onboard IMUs, the device could restrict beeping to instances where physical movement is detected---a lost device normally stays static, and its movement could indicate malicious tracking.}
    
    \item \blue{\textbf{Faster change of MAC address.} Accelerating the MAC address rotation frequency requires attackers to lock the identity with fewer RSSI samples per MAC address, affecting clustering accuracy and prolonging the attacker's navigation time and distance.}
    
    \item \blue{\textbf{Unfixed BLE advertising power and interval.} Varying the BLE advertising power and transmission intervals introduces artificial signal variance. This renders the attacker's spatial (RSSI) and temporal measurements fundamentally unstable, effectively defeating RSSI-IMU navigation.}
\end{enumerate}

\blue{The three countermeasures are orthogonal and can be combined to provide a highly secure Find My ecosystem.}

\section{Open Science}

\blue{To mitigate potential misuse, we have released a limited version of the Snatcher framework. The \textbf{SnatchAPP} Android application is publicly available with only the \textit{lost device discovery} and \textit{non-owner sound trigger} functionalities, while the navigation assistance modules have been deliberately omitted. The ESP32 firmware for research evaluation is fully released. 
The artifact is available at \url{https://github.com/rzy0901/Snatcher}, where the repository is organized into three subdirectories:}

\begin{itemize}
    \item \blue{\textbf{SnatchAPP}: An Android implementation for discovering lost Find My devices and triggering non-owner sound alerts. Navigation assistance capabilities are excluded to prevent misuse.}
    \item \textbf{esp32\_airtag\_mac\_rotation}: ESP32 firmware that emulates the rapid MAC address rotation behavior of an AirTag.
    \item \textbf{esp32\_sound\_maker}: ESP32 firmware to remotely trigger sounds on nearby, non-owner Apple accessories.
\end{itemize}

\end{document}